\documentclass[sigconf]{acmart}
\AtBeginDocument{%
  }

\setcopyright{acmlicensed}
\copyrightyear{2025}
\acmYear{2025}
\acmDOI{10.1145/3706598.3714097}

\acmConference[CHI '25]{ACM Conference on Human Factors in Computing Systems}{April 26--May 01, 2025}{Yokohama, Japan}
\acmISBN{979-8-4007-1394-1/25/04}

\usepackage{subcaption}
\usepackage{enumitem}
\usepackage{multicol}
\usepackage{xcolor} 

\newcommand{\edit}[1]{\textcolor{black}{#1}}

\newcommand{\control}{\textbf{Control}}
\newcommand{\implicit}{\textbf{Implicit Answer}}
\newcommand{\hl}{\textbf{Uncertainty Highlighting}}
\newcommand{\discl}{\textbf{Reliance Disclaimer}}
\usepackage{multirow}
\newcommand{\xhdr}[1]{\vspace{1.7mm}\noindent{{\bf \textit{#1.}}}}
\begin{document}

\title[To Rely or Not to Rely?]{\edit{To Rely or Not to Rely?} Evaluating Interventions for \\ Appropriate Reliance on Large Language Models}

\author{Jessica Y. Bo}
\email{jbo@cs.toronto.edu}
\affiliation{%
  \institution{University of Toronto}
  \city{Toronto}
  \country{Canada}
}

\author{Sophia Wan}
\email{sophia.wan@alumni.utoronto.ca}
\affiliation{%
  \institution{University of Toronto}
  \city{Toronto}
  \country{Canada}
}

\author{Ashton Anderson}
\email{ashton@cs.toronto.edu}
\affiliation{%
  \institution{University of Toronto}
  \city{Toronto}
  \country{Canada}
}

\renewcommand{\shortauthors}{Bo, Wan, and Anderson}

\begin{abstract}
    As Large Language Models become integral to decision-making, optimism about their power is tempered with concern over their errors. Users may over-rely on LLM advice that is confidently stated but wrong, or under-rely due to mistrust. Reliance interventions have been developed to help users of LLMs, but they lack rigorous evaluation for \edit{appropriate reliance}. We \edit{benchmark the performance of three relevant interventions} by conducting a randomized online experiment with 400 participants attempting two challenging tasks: LSAT logical reasoning and image-based numerical estimation. For each question, participants first answered independently, then received LLM advice modified by one of three reliance interventions and answered the question again. Our findings indicate that while interventions reduce over-reliance, they generally fail to improve appropriate reliance. \edit{Furthermore, people became more confident after making incorrect reliance decisions in certain contexts, demonstrating poor calibration. } Based on our findings, we discuss implications for designing effective reliance interventions in human-LLM collaboration.
\end{abstract}


\begin{teaserfigure}
\centering
  \includegraphics[width=0.9\textwidth]{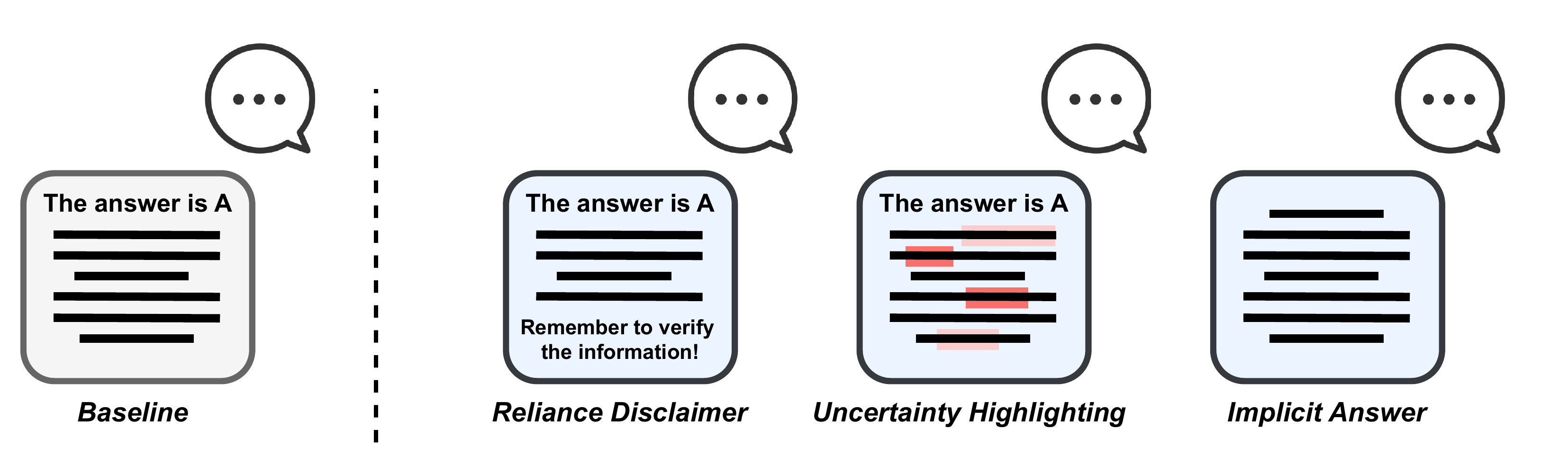}
  \caption{LLM reliance interventions benchmarked: \discl, \hl, and \implicit.}
  \label{fig:intervention}
\end{teaserfigure}

\received{20 February 2007}
\received[revised]{12 March 2009}
\received[accepted]{5 June 2009}

\maketitle

\section{Introduction}
\label{sec:intro}

As people increasingly incorporate Large Language Models (LLMs) into their decision-making workflows, there is cause for both tremendous optimism and skeptical concern \cite{zhao2023survey, eigner2024determinants}. LLMs encode remarkably vast amounts of knowledge; can communicate with users in clear, natural language; and have the potential to improve both the quality and efficiency of users' decision-making \cite{tankelevitch2024metacognitive}. However, people can over-rely on LLM advice \cite{spatharioti2023comparing, kabir2024stack, si2023large} or, in contrast, under-rely on it if they mistrust their LLM partner \cite{zhang2023taking}. LLMs are prone to hallucination and error, yet often present their outputs with impressive detail and confidence \cite{zhang2023siren, chen2023combating}. In contrast, algorithm aversion---the tendency to distrust algorithmic agents---can lead users to ignore helpful recommendations due to general distrust towards the technology \cite{dietvorst2015algorithm, burton2020systematic}.

Conscious effort must therefore be placed on helping users calibrate appropriate levels of trust and reliance on LLM advice \cite{zhang2020effect, zhou2024rel}. Neither over-reliance nor under-reliance is ideal; instead, the focus should be to achieve appropriate reliance \cite{schemmer2022should, ma2023should, he2023knowing}. 
Research in the tangential field of human-AI collaboration has explored methods like explanations and calibration techniques to improve joint decision-making by enhancing AI transparency and verifiability, though results often show mixed outcomes and trade-offs \cite{vasconcelos2023explanations, liao2021human, miller2019explanation, schoeffer2024explanations, bansal2021does}. \edit{In this work, we adapt rigorous measures of appropriate reliance from prior works in human-AI collaboration \cite{schemmer2023appropriate} to benchmark the impact of several LLM reliance interventions in human-LLM collaboration.}

\edit{Due to the risk of hallucinations, reliance and calibration interventions in LLMs are often motivated by curbing over-relying on wrongful information \cite{kim2024m, si2023large, spatharioti2023comparing}, however recent attention has also shifted to optimizing for appropriate reliance \cite{zhou2024rel, zhang2023taking, collins2024modulating, do2024facilitating}. }
Reliance interventions that involve communicating the capabilities and confidence of an LLM to its users \cite{spatharioti2023comparing}, or introducing various types of frictions that the user must bypass to receive the advice \cite{collins2024modulating}, are being actively developed and evaluated. Since interventions are evolving rapidly, many of these methods still lack comprehensive evaluations \edit{of appropriate reliance in real users, especially those focusing on technical contributions \cite{lin2022teaching, becker2024cycles, bowman2022measuring}}.
\edit{Furthermore, people's subjective perceptions---such as their self-confidence---can have a complex relationship with their reliance behaviour  \cite{kim2024m, cao2022understanding, lu2021human, si2023large}, which warrants further exploration. 
Finally, the impact of different intervention types on reliance and perceptions remains unclear, as existing studies do not benchmark multiple interventions.}
We seek to address these gaps by asking the following research questions:
\begin{itemize}[topsep=4pt]
\item \textbf{RQ1}: How do different reliance interventions affect users' tendencies to over- and under-rely on LLM advice?
\item \textbf{RQ2}: Can reliance interventions appropriately calibrate users' confidence in LLM advice?
\end{itemize}

\xhdr{Summary of Contributions} We conduct a pre-registered, randomized online experiment with $n=400$ participants where they attempted two diverse and challenging question-answering tasks: LSAT logical reasoning and image-based numerical estimation. Participants first attempted the tasks on their own, then were shown LLM advice that was randomly treated with one of four conditions (three reliance interventions and one baseline) and given the question again. To evaluate reliance, we implemented a rigorous set of evaluation metrics that isolate the decision-making process users undergo when choosing to rely on the LLM's advice or their own judgment. 

We find that while interventions can significantly reduce over-reliance, they are generally \textit{not effective at improving appropriate reliance}. These interventions tend to reduce reliance overall---reducing over-reliance at the expense of useful reliance. Participants were also generally miscalibrated in their confidence, reporting a higher increase in confidence when they \textit{make the wrong reliance decision}. Out of all the interventions, \discl \space is the only one to improve appropriate reliance and better calibrate confidence when participants relied on LLM advice, but this effect was only observed on the LSAT task. \hl \space worsened subjective perceptions and had the least effect on reliance out of the three interventions. \implicit \space induced better self-reliance in both tasks, but at the cost of requiring significantly more time for participants to process the information. We also discuss the implications of our results for the design and evaluation of appropriate reliance interventions for the future of human-LLM interactions. We highlight the need to perform rigorous, human-centered evaluations of technical LLM calibration methods, particularly with emphasis on achieving appropriate reliance. Furthermore, as the trade-off between balancing over- and under-reliance may be difficult to avoid, defining which outcome to prioritize should be done within the context of the specific human-LLM collaboration task.

\section{Related Works}
\label{sec:related}

We provide an overview of the growing body of research in human-AI interactions, human-LLM collaboration, and LLM calibration techniques that have been explored in both human-computer interaction (HCI) and AI spaces. 

\subsection{Appropriate Reliance on AI}

The need for appropriate reliance has been a constant over the years of technological development \cite{lee2004trust}. With the advancement of AI technologies, much attention has turned on how end users can incorporate assistive AI models into their decision-making tasks \cite{lai2021towards, kleinberg2018human, chen2023understanding}. Given that AI predictions are not guaranteed to be accurate, the user must decide whether to or not to incorporate the AI's advice \cite{sohrawardi2024dungeons, chiang2022exploring, nguyen2018believe, inkpen2023advancing, lu2021human, cao2022understanding}. The concept of appropriate reliance can typically be defined as \textit{"relying on the AI when it's correct, and relying on yourself when it's not"}---which concerns reducing both over-reliance and under-reliance \cite{schemmer2023appropriate, schoeffer2024explanations}. Root causes for inappropriate reliance can come from misreliance (inherently nonoptimal rate of reliance) or discrimination errors (being unable to identify the payoff of following the AI relative to following oneself) \cite{guo2024decision, ma2023should}.

Factors that influence reliance include the expertise of the user in the task, their cognitive engagement in the task, and the information presented by the AI \cite{chen2023understanding, he2023knowing, gajos2022people}. On the individual level, people may also be averse or appreciative towards algorithms, perhaps through pre-conceptions or judgments formed after seeing the AI's performance \cite{kim2023algorithms, dietvorst2015algorithm, lu2023strategic, nourani2021anchoring, ma2023should}. \edit{A number of strategies have been developed to augment users' engagement and to provide them with more information of the AI decision, such as through providing explanations \cite{reingold2024dissenting, vasconcelos2023explanations, zhang2020effect, schoeffer2024explanations}.}
The best interventions for appropriate reliance should be designed to support the users' abilities to \textit{verify} the AI's recommendation, not necessarily to provide a fully transparent system \cite{fok2023search, poursabzi2021manipulating, wang2021explanations}. 
\edit{We adapt appropriate reliance metrics \cite{schemmer2023appropriate} to assess the ability of LLM interventions to help users verify outputs and increase appropriate reliance}.

\subsection{LLMs as Decision Aids}

Recent research has investigated the role of LLMs in human-LLM collaborations for applications such as writing \cite{liang2024monitoring}, communication \cite{fu2024text}, coding \cite{kazemitabaar2023studying, nam2024using}, education \cite{lo2023impact, kumar2023math, bo2024disclosures}, fact-checking \cite{wolfe2024impact, kim2024can, si2023large}, information search \cite{spatharioti2023comparing, salikutluk2023interacting, sharma2024generative}, and other specialized domains \cite{choudhury2023investigating, cheong2024not, yang2023large}. In these tasks, the LLM can provide guidance and information for the user to assist in their decision-making, learning, and creations \cite{zhang2023taking}. They can also offer antagonistic advice to induce the user into considering alternative options \cite{chiang2024enhancing, ma2024towards, cai2024antagonistic}. Even when the user is just reading without explicitly searching for advice, the opinions embedded in the LLM outputs can be influence the user's own opinions \cite{ferguson2023something, sharma2024generative, krugel2023chatgpt}.
In technical AI research, the quality of LLM responses has been improved through methods such as chain-of-thought with intermediate reasoning steps \cite{wei2022chain}, debating amongst multiple LLMs \cite{khan2024debating, kim2024can}, and enforcing self-consistency in outputs \cite{wang2022self, becker2024cycles}.

In many cases, the human-LLM team can reach better performance as a team than individually, tapping into the strength of their combined knowledge and skill sets \cite{bowman2022measuring}.  However, the question of how to induce \textit{appropriate reliance} is still open. While using an LLM for information search can be just as efficient as a search engine, the user's ability to distinguish incorrect information is diminished, resulting in a higher rate of over-reliance on incorrect information \cite{si2023large, xu2023chatgpt}. On the other hand, aversion to LLM advice has also been observed, resulting in users losing performance gains that they could have achieved through relying on the LLM \cite{zhang2023taking}.

\edit{In addition to technical limitations of LLMs, human decision-makers are also imperfect and subject to cognitive biases  \cite{bertrand2022cognitive}. People may have unrealistic expectations of LLMs and use them inappropriately \cite{vafa2024large}, such as by assuming LLMs are skilled at math simply because they are computational models \cite{zhou2024rel}. People also demonstrate divergent decision-making characteristics, with some preferring to meticulously optimize their decision while others take the satisficing or minimizing paths that allow them to bypass extensive thinking \cite{kaur2024interpretability, misuraca2015decision}. Those who minimize time spent on tasks may be more likely to over-rely on LLM advice, given its perceived completeness and relevance \cite{eigner2024determinants}.}

\subsection{Uncertainty and Reliance Calibration Techniques}

Much attention has been placed on promoting transparency in LLMs, with the goal of communicating their capabilities or confidence to users \cite{liao2023ai, yang2024report, lin2022teaching}. 
For example, \hl \space draws visual attention to uncertain tokens in the LLM's outputs that the user may want to scrutinize \cite{do2024facilitating, vasconcelos2023generation},
which have been found to reduce over-reliance in information search \cite{spatharioti2023comparing}, improve trust in answering questions \cite{do2024facilitating}, and increase interpretability in AI-aided coding \cite{vasconcelos2023generation}. Other interventions that have been tested in human-LLM collaborations include contrastive explanations \cite{si2023large}, adding semantic expressions of uncertainty \cite{kim2024m, lin2022teaching, steyvers2024calibration}, and confidence scores \cite{virk2024enhancing}. 

Complex LLM calibrations may carry the risk of being too cognitively overwhelming to understand---for example, to truly understand uncertainty highlighting, users must know that the highlighting reflects token-wise uncertainty rather than semantic uncertainty. Many people make satisficing decisions, where they only look at information to make `good enough' decisions but don't engage further, thus rendering explanations and interventions ineffective \cite{kaur2024interpretability, eigner2024determinants}. \discl \space is a low-effort intervention that can be statically applied to LLM outputs, such as through a persistent warning on the chat interface that encourages the user to verify the LLM's answers. While extremely simple, it has been found to reduce students' over-reliance on incorrect LLM guidance in a classroom setting \cite{bo2024disclosures}. 

Additionally, explainable AI (XAI) research has found it effective to focus on increasing the user's cognitive engagement in the task, such as through simple cognitive forcing functions that add frictions and delays to the interface or removing direct recommendations \cite{buccinca2021trust, gajos2022people}. Similar friction-based UI designs have been applied to the LLM space to modulate user behaviour \cite{collins2024modulating}. 
We take inspiration from the domain of LLMs for education, where LLMs are prompted to provide guidance or hints instead of direct answers \cite{kumar2023impact, ma2024teach}. Friction is introduced in the reliance process, where users have to cognitively engage to consider the recommended answer.

In this study, we implement three highly relevant interventions \edit{that induce} cognitive engagement \edit{in different ways}: \discl, \hl, and \implicit. 

\section{Methodology}
\label{sec:method}
We designed a large-scale, randomized, and incentivized experiment to evaluate the efficacy of three reliance interventions applied to the outputs of GPT-4o\footnote{\edit{We selected GPT-4o (https://openai.com/index/hello-gpt-4o/) as it is the most state-of-art LLM at the time of experimentation.}} in two question-answering tasks. 
To answer our research questions, we will compare the effect of the interventions on performance, reliance behaviour, confidence calibration, and time of engagement against the baseline LLM advice in the \control \space condition. In this section, we outline our choices of tasks, procedure, measures, and conditions. 


\subsection{Question-Answering Tasks}
We gave the participants two separate question-answering tasks, which we carefully selected to meet the following desiderata: 
\begin{itemize}[noitemsep,topsep=0pt]
    \item [D1]\textit{\textbf{Accessible to a general audience.} The tasks should not require any specialized knowledge or skills, so that they are relevant and accessible to a typical person.} 
    \item [D2]\textit{\textbf{Difficult.} The tasks should be sufficiently challenging, requiring a meaningful level of cognitive effort to answer correctly.}
    \item [D3]\textit{\textbf{Resistant to cheating.} The tasks should not be straightforward to solve by using external assistance, such as a search engine.}
\end{itemize}

Driven by the desiderata above, one of our central objectives was to select tasks that a typical person could reason about, even if they do not arrive at the correct answer. We therefore intentionally excluded factual questions characterized by domain-specific knowledge that individuals either possess or lack, and cannot conclude through reasoning alone. This approach emphasizes critical thinking over recall of factual information, and ensures that the average person has the potential to get the correct answer, as well as the ability to engage with LLM advice and reliance interventions. To increase generalizability, we selected two different question formats: multiple-choice, a common standard in human-AI collaboration research; and numerical estimation, which allows for a more granular measurement of the LLM's influence on participants through the weight-of-advice (WoA). Below, we outline the tasks selected for each format.

\xhdr{LSAT Logical Reasoning}
Based on preliminary results from piloting multiple-choice questions from several domains, we found that logical reasoning questions best balanced difficulty with accessibility. We selected LSAT logical reasoning questions as the domain for the multiple-choice task and adapted two multiple-choice LSAT logical reasoning questions, both labeled `difficult', for the first task. Each LSAT question has five options, of which exactly one is correct. We generated both correct and incorrect LLM advice for each question for use during the experiment. The questions and some sample LLM advice can be found in Figures \ref{fig:lr_q_road}-\ref{fig:lr_q_star} in Appendix~\ref{app:questions}. \edit{Correct answers were determined using LSAT answer keys.}

While LSAT questions are highly detailed, they are designed to test for aptitude in law school rather than knowledge of the legal domain. As such, they do not contain undefined jargon, nor do they require domain knowledge (satisfying \textit{D1}). In addition to reasoning, LSAT Logical Reasoning questions also test analytical and reading comprehension skills, making them non-trivial to answer (satisfying \textit{D2}). Furthermore, we reworded a question so that entering it into a search engine would not result in the correct answer (satisfying \textit{D3}). Finally, they lend our study further external validity by emulating situations where people use LLMs to assist them with text comprehension or reasoning, in addition to people studying for tests such as the LSAT. While the MMLU dataset is another popular multiple-choice question benchmark for LLMs \cite{hendrycks2020measuring},  we did not find it suitable since it is skewed towards fact-based, \textit{know-it-or-you-don't} questions. 

\xhdr{Numerical Estimation}
While most previous studies in human-AI collaboration focus on multiple-choice or binary outcome questions, we believe that introducing a numerical estimation task adds a valuable new dimension to this research space. Among the various forms of numerical estimation, such as predicting values in regression tasks, we opted for an image-based format based on a classic carnival game: estimating the quantity of beans in a glass jar. We posit that this is a gamified task that is familiar to most people and does not require advanced mathematics (satisfying \textit{D1}). Despite the simplicity of the estimation process, it is still challenging to achieve accurate results
(satisfying \textit{D2}). Lastly, since we took the pictures and many of the beans are occluded, it is impossible to use external assistance to ``cheat'' on this task (satisfying \textit{D3}).

For the experiment, we captured photos of glass jars with \edit{manually-counted} quantities of beans from both side perspectives and bottom perspectives. One set featured a smaller jar containing 403 chickpeas, while the other contained a larger jar with 856 jellybeans. 
The pictures and sample LLM advice can be found in Figures~\ref{fig:ne_q_chickpea}-\ref{fig:ne_q_jellybean} in Appendix~\ref{app:questions}.
Unlike in multiple-choice tasks, where advice can be objectively classified as good or bad, the goodness of the advice in the numerical estimation setting depends on the participant’s initial estimate. \edit{For example, while advice recommending a number that is far lower than the ground truth is objectively wrong, it is still \textit{better} than an even lower participant estimate. }To account for this variability, we provided five distinct levels of LLM advice relative to the ground truth quantity: \textit{very low}, \textit{low}, \textit{good}, \textit{high}, and \textit{very high}. \edit{We chose these levels of advice based on results from initial pilot experiments to balance the proportion of bad advice with good advice. Visualizations of the advice levels with the distribution of participant guesses are shown in Figure \ref{fig:five_levels} in Appendix \ref{app:five_levels}.}

\begin{figure*}[t!]
    \centering
    \begin{subfigure}[t]{\linewidth}
        \centering
        \includegraphics[width=\linewidth]{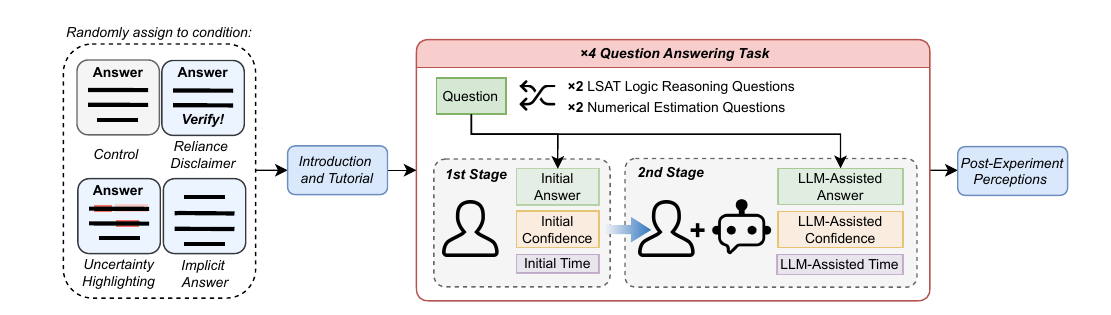}
        \caption{Diagram of the full experimental procedure.}
        \label{fig:procedure}
    \end{subfigure}%
    \\
    \begin{subfigure}[t]{\linewidth}
        \centering
        \includegraphics[width=0.92\linewidth]{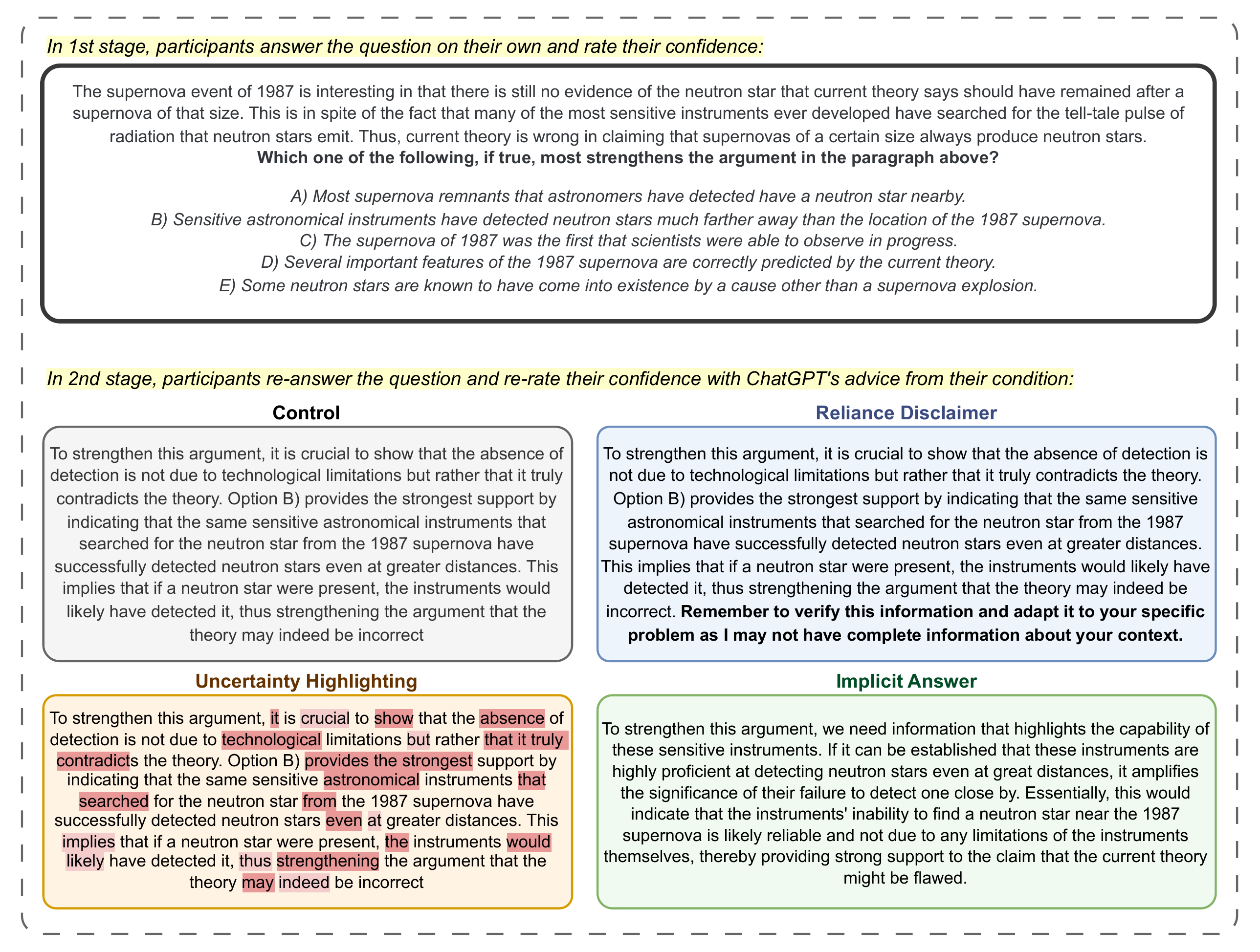}
        \caption{\edit{Sample question and advice provided from the LSAT task (more examples of the questions and advice are in Appendices \ref{app:questions} and \ref{app:advice}). Participants saw only one out of the four advice based on their condition.}}
        \label{fig:questions}
    \end{subfigure}
    \caption{\edit{Diagrams and materials related to the procedure and tasks, where (a) shows the experimental flow, and (b) shows a complete question from the LSAT task and the advice from all four conditions that participants were provided. }}
    \label{fig:procedure_questions}
    \vspace{3em}
\end{figure*}

\subsection{Procedure}

In the experiment, each participant was randomly assigned to one of four conditions and then answered two questions from the LSAT task and two questions from the Numerical Estimation task, randomized so that the LSAT phase and Numerical Estimation phase were equally likely to appear first. 
The orders of questions within each category were also randomized for each participant. Participants received pre-generated advice from GPT-4o. This advice was randomly sampled for each question, meaning participants could receive any mix of good or bad advice. After completing the tasks, participants completed a short post-survey. See Figure~\ref{fig:procedure} for the full diagram of the experimental flow and Figure~\ref{fig:questions} for sample advice from all four conditions. 

To mimic the stakes and level of engagement typically involved in real-world decision-making scenarios where people might consult AI chatbots, we incorporated two key design elements into the study: 
\begin{enumerate}[topsep=4pt]
    \item \textbf{Two-Stage Answering:}  \edit{We aim to emulate contexts where the user is already engaged in the task before turning to LLM advice, such as a student who tries to solve the problem on their own first before checking with ChatGPT. Given that our participants are crowdworkers without prior context or investment in the tasks, we believed it is necessary to induce engagement.}  To address this, we require participants to answer each question on their own before seeing the LLM-generated advice, and then once again after they are given the advice. \edit{Studies that use one-stage questioning, where the LLM’s advice is presented alongside the initial question, reflect a context where the participant has less incentive to be engaged, which can lead to artificially high over-reliance rates. }

    \item \textbf{Incentivization:} Real-world decision-making involves balancing risk and reward. To simulate these stakes and induce incentive-compatible behaviour, we introduced a bonus, amounting to a maximum of 20\% of the total participation pay. Participants earn rewards for correct answers, with two opportunities for bonus pay: one before, and one after receiving LLM-generated advice. In the LSAT questions, participants received a bonus if they selected the correct answer, while for the numerical estimation task, participants earned a bonus if they guessed within 10\% of the true count.
\end{enumerate}

\subsection{Measures}
\label{sec:equations}
Our research questions focus on reliance behaviour (\textbf{RQ1}) and confidence calibration (\textbf{RQ2}). We detail the equations for these categories of measurements, in addition to secondary measurements of task performance and time of engagement.

\begin{figure}
    \centering
    \includegraphics[width=\linewidth]{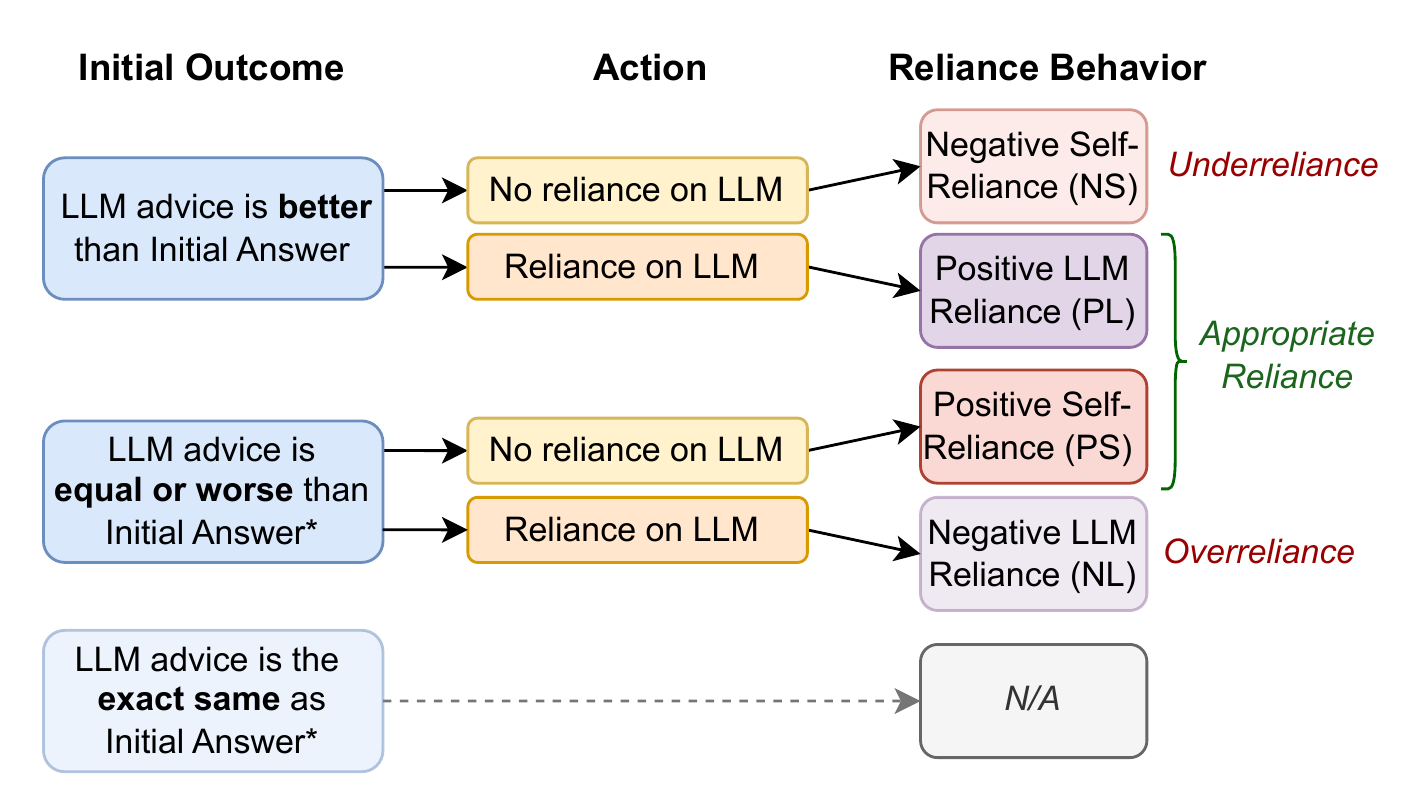}
    \caption{Illustration of the reliance pathways based on the initial answer and LLM advice. Note that in the second condition, \textbf{equal or worse} refers to cases where the LLM advice is \textit{different} from the initial answer, but the relative accuracy can be equal (such as both answering a multiple-choice question wrong). Correspondingly, the third condition only covers cases where the LLM advice matches the initial answer. }
    \label{fig:pathways}
\end{figure}

\xhdr{Reliance}
Following existing literature on appropriate reliance in human-AI interactions, we use the two-stage question-answering task format where a participant first answers a given question on their own before receiving LLM advice and answering again. 
If the participant's initial answer conflicts with the LLM's advice, appropriate reliance requires the human to correctly determine whether the LLM's advice is better or worse. 
We focus on this explicit \textit{decision-making step} as the construct of interest for measuring reliance, where the user is presented with the choice to either a) change their answer to rely on the LLM's advice, or b) disregard it and retain their initial answer.
Thus in our analysis on reliance, we \textit{only consider the data instances} where the participant's initial answer differs from the LLM's advice. See Figure \ref{fig:pathways} for an illustration of the reliance decisions pathways. 

What does it mean to \textit{rely} on LLM advice? In multiple-choice questions, we follow previous work by defining reliance as \textit{changing one's answer to the LLM's suggestion}. In estimation questions, we define reliance as \textit{an answer with positive weight-of-advice (WoA)},
where WoA is calculated as $ \frac{Final\ Estimate\ -\ Initial\ Estimate}{LLM\ Advice\ -\ Initial\ Estimate}$. We further define the reliance outcomes as falling into one of four categories: 
\begin{itemize}[noitemsep,topsep=2pt]
    \item \textit{Positive LLM Reliance (PL)}:  the participant changes their original answer to follow the LLM’s advice, and subsequently improves their performance.
    \item \textit{Positive Self-Reliance (PS)}: the participant does not follow the LLM’s advice, and doing so would have worsened \edit{or failed to improve} their performance.
    \item \textit{Negative LLM Reliance (NL)}: the participant changes their original answer to follow the LLM’s advice, but \edit{worsens or fails to improve their performance}. 
    \item \textit{Negative Self-Reliance (NS)}: the participant does not follow the LLM’s advice, but doing so would have improved their performance.
\end{itemize}

PL and PS can be grouped together as \textit{Correct Reliance Behaviours}, while NL and NS are \textit{Wrong Reliance Behaviours}. NL represents over-reliance on LLM advice (the participant took the advice when it \edit{worsens or fails to improve improve} their performance) and NS represents under-reliance (the participant didn't take the advice but should have). 
\edit{We note that PL is possible even if the LLM advice is incorrect, as long as there is a notion of relative error---such as switching to a \textit{less} wrong answer suggested by the LLM, which applies to the Numerical Estimation task and can extend to other real-world tasks. In contrast, the LSAT task lacks relative scoring across options, treating all incorrect answers as equally wrong; so relying on incorrect advice counts as over-reliance. This disambiguation is made since the LLM advice would not improve performance, so it would be preferable for the user to engage in the task and self-rely, allowing them to learn from the experience \cite{gajos2022people}.}
From these behaviour classifications, we compute the following \textit{appropriate reliance} metrics: 
\begin{itemize}[itemsep=4pt,topsep=4pt]    
    \item \textbf{Relative LLM Reliance (RLR)} $= \frac{PL}{PL + NS}$ = Rate of appropriate reliance on correct LLM advice. 
    \item \textbf{Relative Self-Reliance (RSR)}  $= \frac{PS}{PS + NL}$ = Rate of appropriate self-reliance when the LLM advice is wrong. 
    \item \textbf{Appropriate Reliance Ratio (ARR)}  $= \frac{PL \ / \ (PL + NS)}{NL /\ (PS + NL)}$  = Ratio of appropriate reliance on correct advice to over-reliance on wrong advice, where:
    \begin{itemize}[noitemsep,topsep=0pt]
        \item \textit{ARR < 1}: Higher rate of over-reliance than positive reliance. 
        \item \textit{ARR = 1}: Random chance; LLM reliance rates are the same regardless of LLM correctness.
        \item \textit{ARR > 1}: Higher rate of positive reliance than over-reliance. 
    \end{itemize}
\end{itemize}

We compute the metrics for each condition across participants, then calculate the effect size as the difference in outcomes between the control and each intervention. To assess statistical significance, we use permutation tests. For each planned comparison, we generate a distribution of effect sizes under the null hypothesis by randomly shuffling and reassigning data between the control and intervention groups. By comparing the observed effect size to this shuffled distribution, we estimate how likely it is to occur by chance, using a significance threshold of $p<.05$.

\xhdr{Confidence and Subjective Perceptions}
For each question, we collect \edit{the participant's confidence in the answer they gave prior to receiving LLM advice (\textbf{Initial Confidence}), and their confidence in the answer they gave after (\textbf{Final Confidence}) receiving the LLM-generated advice}. Confidence is measured using a \edit{unipolar Likert} scale from 1 (highly uncertain) to 5 (highly certain), which allows us to calculate the \textbf{Change in Confidence} from the first to second stage. Well-calibrated confidence should result in lower confidence scores when participants make incorrect reliance decisions. We compare the change in confidence between correct and wrong reliance outcomes for each of the conditions, where statistical differences are tested with Mann-Whitney U-test for non-normal distributions with significance threshold $p<.05$.

At the end of the survey, participants provide additional ratings on a \edit{bipolar} Likert scale of 1 (disagree) to 7 (agree) for the following subjective perception statements:
\begin{itemize}[noitemsep,topsep=0pt]
    \item \textit{ChatGPT's advice made me more confident.\footnote{\edit{This subjective confidence measure is separate from the question-level confidence measure (the actual target of RQ2).}}}
    \item \textit{ChatGPT's advice seems accurate.}
    \item \textit{ChatGPT's advice was helpful.}
\end{itemize}

\xhdr{Task Performance}
Since our experiment is not designed to emulate the realistic performance of LLMs in tasks, we do not particularly emphasize the human-LLM collaborative task performance. However, we explore the initial error from participants independently answering questions, and compare this with their updated error after re-answering questions with LLM advice. We also compute a \textbf{Best Possible Performance} metric, which simulates the outcomes as if the human made the correct reliance decision in every single question, representing the best possible accuracy of a human-LLM team.

\xhdr{Time of Engagement}
To understand participant engagement with the LLM's advice, we measure the amount of time that participants spend on answering each stage of question-answering. From these two time measurements, we calculate the relative time as the ratio between time taken to engage with the advice, and the time taken to answer the initial question (\textbf{Relative Advice Time}).

\subsection{Experimental Conditions}
In this section, we outline the implementation of our four experimental conditions, each representing different forms of LLM-generated advice. The interventions were chosen based on their relevance to the research community, their applicability across both multiple-choice and estimation tasks, and their diversity in design---incorporating both static and dynamic interventions, and both visual and semantic augmentations. Whenever possible, we used direct outputs generated through tailored prompts from the LLM with minimal manual adjustments. Figure 1 provides a visual overview of the advice conditions, with more details available in Figure \ref{fig:questions}.

\xhdr{Control}
To generate the baseline advice, we input the questions into GPT-4o through OpenAI's API and record the responses directly. Participants receive a direct recommendation of the answer (such as ``Option A" in Logical Reasoning and ``120 chickpeas" in Numerical Analysis) in addition to an explanation supporting the answer. 

\xhdr{Reliance Disclaimer}
The following static disclaimer statement is appended to all advice generated in \control: \textit{``Remember to verify this information and adapt it to your specific problem as I may not have complete information about your context"}. This text was generated through prompting the LLM to disclose its limitation as a language model to convey that its outputs should be used carefully. For consistency and to emulate real-world chatbot disclaimers like ChatGPT, we apply the same disclaimer to all advice. \edit{However, we acknowledge that the effectiveness of a static disclaimer may not persist in long-term deployment. Alarm fatigue describes the process of desensitization, where users learn to ignore unfounded warnings over time  \cite{cvach2012monitor}. We hypothesize that disclaimers may be most effective for priming new users or in shorter interactions, like a tutoring chatbot that students only use occasionally. Some alternative ways to improve longevity of the effect are to dynamically insert disclaimers fluidly into the conversation---which makes them less repetitive and predictable \cite{bo2024disclosures}, or to insert them only when error rates are particularly high \cite{cvach2012monitor}.} 

\xhdr{Uncertainty Highlighting}
When generating the \control \space advice, we record the log probability of each token. \edit{Our choice to highlight generation probability follows prior works like \cite{spatharioti2023comparing}, but we note that alternative methods of highlighting can offer better interpretation, such as edit probability \cite{vasconcelos2023generation}.} To visually augment the baseline advice, we highlight the tokens with log probability under -1 in \colorbox{purple}{stronger red} and indicate them to be \textit{uncertain}, and the tokens under -0.5 in  \colorbox{pink}{lighter pink} and indicate them to be \textit{slightly uncertain}. We selected these thresholds to balance highlighting uncertain text within the answer with minimizing visual overwhelm for the reader. \edit{We make manual adjustments to maintain coherence, such as extending highlighting to full words if they were only partially highlighted and removing highlighting on articles that do not impact semantic meaning.} Participants were not told that the method of determining uncertainty is based on log probabilities. However, they were reminded of the definition of the highlighting colors in every survey page with the LLM advice.

\xhdr{Implicit Answer} This intervention semantically augments the content of the advice to imply the answer without writing it directly, requiring more engagement to extract the answer that the advice supports.
For the Logical Reasoning questions, we append an instruction to the prompt to never explicitly state the answer's text, and to use paraphrasing to generate an argument in support of the answer. For the Numerical Estimation questions, we manually removed the final calculation step from the advice generated in \control \space (as it was difficult to achieve this through prompting alone). In this way, we ensure that participants see the intermediate steps of the calculation, and can arrive at the value advised by the LLM if they engage with the answer enough to calculate it themselves.

\section{Results}
\label{sec:results}

We pre-registered our experimental design and analysis plan online\footnote{\url{https://aspredicted.org/8BD_WWN}}, then ran the experiment. Here, we report the analyses described in Section \ref{sec:method} separately for LSAT and Numerical Estimation due to the inherent differences between these two tasks. 

\subsection{Participants}

We conducted a power analysis based on preliminary pilot data and found that approximately 200 samples per condition would suffice to achieve significance with an effect size of 0.4 for the ARR. Since we perform the analysis separately for LSAT and Numerical Estimation and each participant answers two questions, we determined that we needed 100 participants per condition. 

We recruited 400 crowdworkers from Prolific, filtering for those who are based in the US, are fluent in English, have a task success rate of 98\% or higher, and have completed at least 50 tasks. Following Prolific's policy, we excluded participants who failed 2/3 attention checks, which was one participant from the \control \space ($n=99$) condition and two from the \hl \space condition ($n=98$). For both of the other conditions, \discl \space and \implicit, all participants passed the attention checks. Across all conditions, the median completion time was 9 minutes and 44 seconds. Participants were compensated \$1.28 USD as the base pay. They had the chance to earn up to \$0.26 USD as a bonus for performance, with the mean earning being \$0.07 USD. Based on Prolific's demographics information about our participants, the participants are 56.3\% female with an average age of 43.5 years and are all located in the US. Furthermore, they are well-distributed in their usage of AI chatbots, with 18.9\% citing daily use, 33.5\% citing weekly use, 17.9\% citing monthly use, and the remaining 29.7\% citing rare or no usage.

\begin{figure*}
    \centering
    \includegraphics[width=0.9\linewidth]{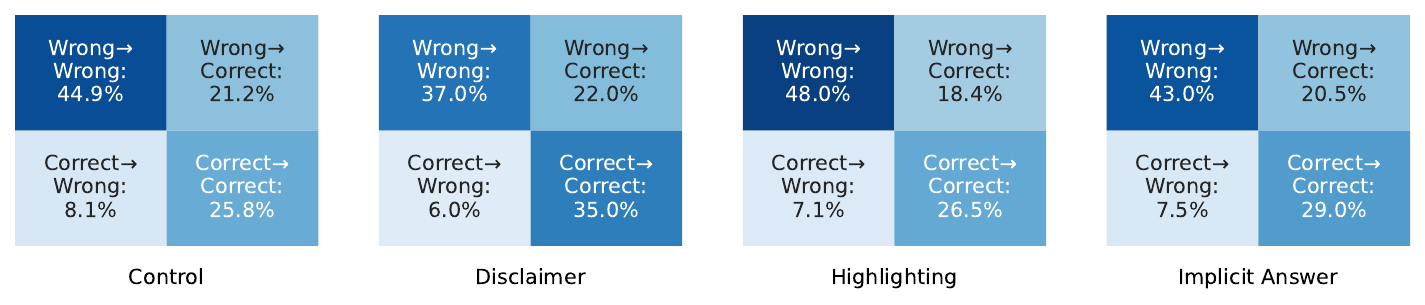}
    \caption{Performance outcomes in the LSAT task for initial and LLM-assisted answers.}
    \label{fig:MC_errors}
\end{figure*}

\begin{figure*}[t!]
    \centering
    \begin{subfigure}[t]{\linewidth}
        \centering
        \includegraphics[width=0.9\linewidth]{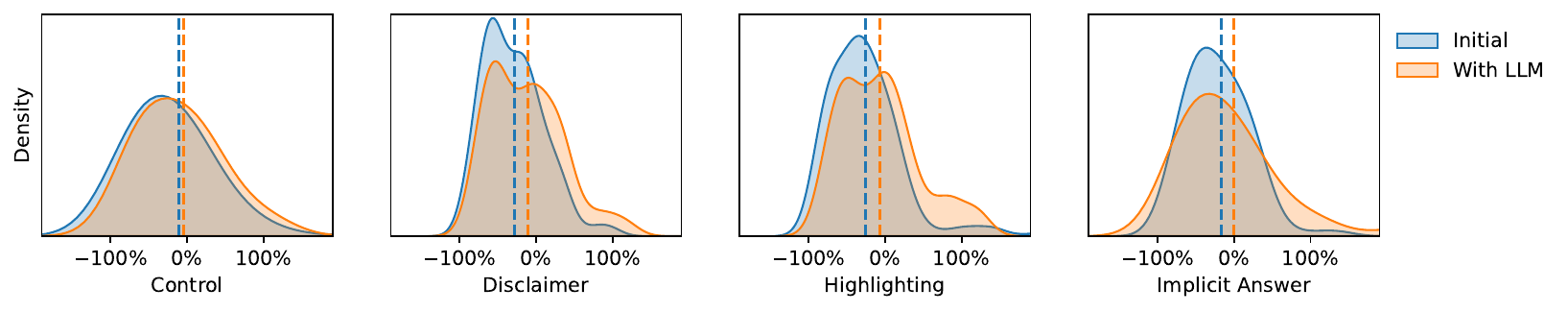}
        \caption{Percent error distributions of the initial and LLM-assisted answers  in Numerical Estimation.}
        \label{fig:NE_errors}
    \end{subfigure}%
    \\
    \begin{subfigure}[t]{\linewidth}
        \centering
        \includegraphics[width=0.9\linewidth]{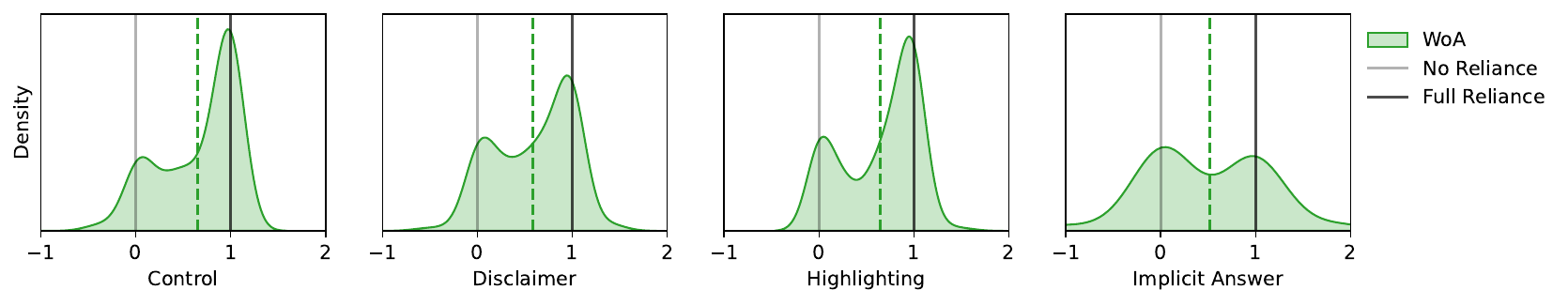}
        \caption{Weight of advice (WoA) distributions in Numerical Estimation.}
        \label{fig:NE_woa}
    \end{subfigure}
    \caption{Performance outcomes in the Numerical Estimation tast for initial and LLM-assisted estimations. (a) shows the percentage error (relative to ground truth) distribution and \edit{(b)} shows the weight-of-advice (WoA) distributions. The mean of each distribution is indicated by the dotted line. }
    \label{fig:NE_performance}
\end{figure*}

\subsection{Task Performance}

We preface our main discussion of reliance with an examination of task performance. We emphasize that since we are interested in how people react to both good and bad LLM advice, we sample these at equal rates. Our LLM performance therefore does not necessarily reflect the actual performance of GPT-4o on our tasks and should not be taken as a measure of realistic human-LLM collaborative decision-making.

\xhdr{LSAT}
Due to the high difficulty of the LSAT questions, participants were initially correct in only 36.0\% of question instances, not far from the rate of random guessing (20\%). Participants then either received correct or wrong LLM advice, equivalent to a 50\% accuracy rate. Their second-stage (LLM-assisted) accuracy rate was 46.1\%. Figure \ref{fig:MC_errors} shows transition matrices of the first- and second-stage performance outcomes in each condition. In the performance outcomes, there are no significant differences between \control \space and any of the interventions as evaluated by a $\chi^2$ test ($p>.05$ for all three interventions). 
Based on the collaborative performance, it is evident that while the LLM improved overall performance, the human-AI team does not perform optimally. In each question, if the human-LLM team relied on the team member that got the correct answer, then the \textbf{Best Possible Performance} would reach 62.1\% accuracy (not 100\%, because sometimes both the human and the LLM answered incorrectly). The gap between optimal and actual performance represents the performance lost to inappropriate reliance behaviour---either relying on wrong LLM advice or ignoring correct LLM advice. 

\xhdr{Numerical Estimation} While estimation may be more approachable, it is still a difficult task. 
The mean \textit{absolute} percentage error relative to the ground truth is 45.7\% across all conditions. With advice, the participants do not significantly improve, answering with a similar \textit{absolute} error rate of 47.0\%---this is unsurprising, since poor LLM advice was given as frequently as good advice. However, in terms of the \textit{mean} (not absolute) percentage error, the participants initially underestimate the counts, averaging -20.2\% relative to the ground truth. With the advice, the gap closes to just -5.6\% (below the ground truth). 
The full distributions, which include both positive and negative percent errors, are shown by condition in Figure \ref{fig:NE_errors}. There are no significant performance differences between conditions (Mann-Whitney U test $p>.05$ for all three interventions). 
The optimal human-LLM team outcome can be simulated by submitting the estimation of the team member who is closest to ground truth. In this hypothetical \textbf{Best Possible Performance}, the mean of the \textit{absolute} error would be 29.0\%. That the observed error was much higher indicates people both over- and under-relied on LLM advice.

\edit{While unrelated to task performance}, we further compute \textit{how much} the participants relied on the advice \edit{for additional insights on their behaviour}. The weight-of-advice taken by each participant in each question is shown as distributions for each condition in Figure \ref{fig:NE_woa}, where a WoA of 1 means fully adopting the LLM's advice, a WoA of 0 means retaining the initial answer, and other values interpolate between these definitions. The distribution of WoA tends to be bimodal, with participants in \control, \discl, and \hl \space predominantly favouring adoption of the LLM advice. Since \implicit \space did not provide the final recommendation, and instead left it to the participants to apply the mathematical hints, the WoA distribution is more flattened, even with values in the negatives or exceeding 1. In our WoA analysis, we drop outlier values over a magnitude of 5 (such as switching from an initial estimate of 600 to a final estimate 4480 when the LLM advice is 544, which constitutes a WoA of -69.3). By the Mann-Whitney U test, \discl's WoA ($p=.03$) and \implicit's WoA ($p=.005$) are significantly smaller than \control's WoA.

\begin{figure*}[t!]
    \centering
    \begin{subfigure}[t]{0.33\linewidth}
        \centering
        \includegraphics[width=\linewidth]{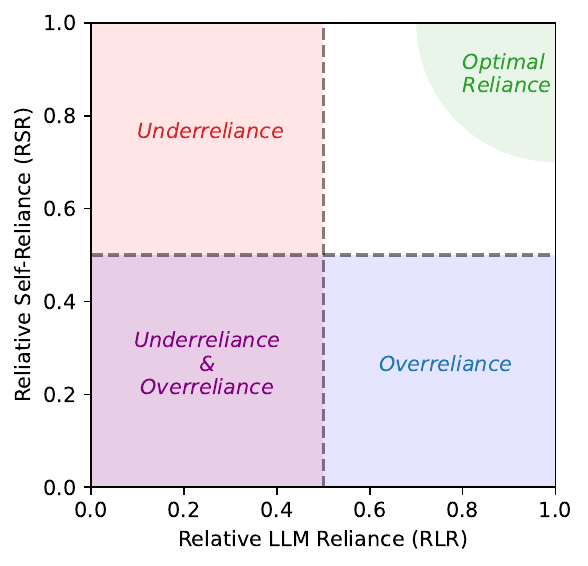}
        \caption{Reference reliance outcomes.}
        \label{fig:reliance_reference}
    \end{subfigure}%
    ~ 
    \begin{subfigure}[t]{0.33\linewidth}
        \centering
        \includegraphics[width=\linewidth]{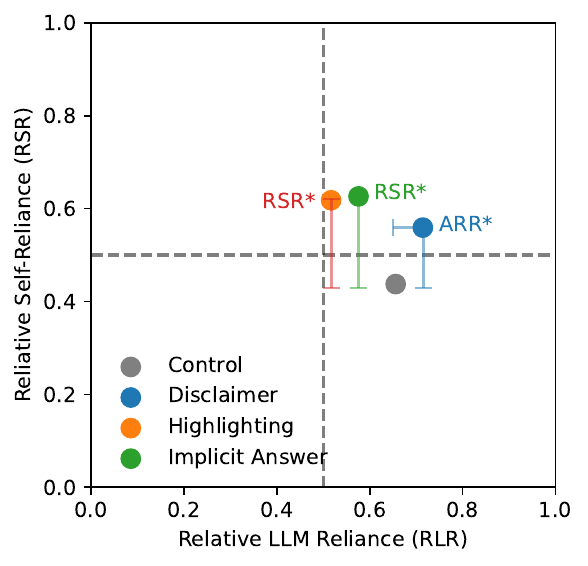}
        \caption{LSAT results.}
        \label{fig:reliance_mc}
    \end{subfigure}
    ~ 
    \begin{subfigure}[t]{0.33\linewidth}
        \centering
        \includegraphics[width=\linewidth]{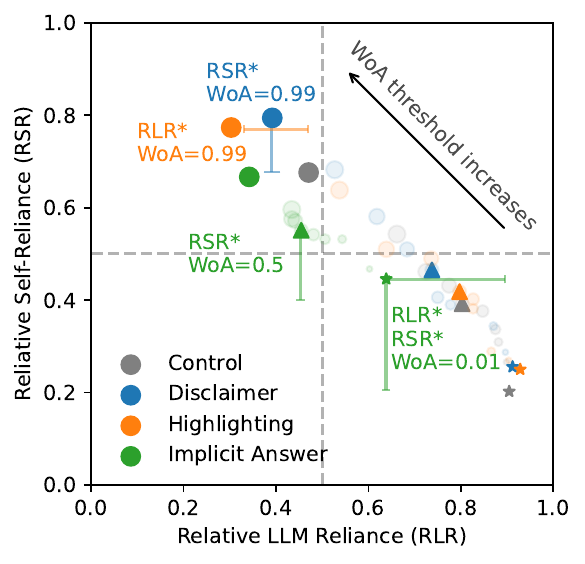}
        \caption{Numerical Estimation results.}
        \label{fig:reliance_ne}
    \end{subfigure}
    \caption{\edit{Visualization of the appropriate reliance measurements -- RLR on the \textit{x-axis}, RSR on the \textit{y-axis}, and ARR indirectly represented as a combination of RLR and RSR (see Section \ref{sec:equations} for equations). Each colored marker represents the \textit{(RSR, RLR)} measurement for a condition. For any significance in the differences between \control \space and each of the interventions, a horizontal bar (RLR), vertical bar (RLR), or both (ARR) are drawn between the two conditions' markers.  Subfigures (a) shows the visual reference for the areas of the graph that correspond to different reliance outcomes.  (b) shows the reliance results for the LSAT task. (c) shows the reliance results for the Numerical Estimation task, using \textit{multiple WoA threshold values} for classifying reliance. Three WoA thresholds are indicated explicitly: 0.01 (star marker), 0.5 (triangle marker), and 0.99 (circle marker), while other thresholds are plotted with translucency to visualize the trajectory of changes.} }
    \label{fig:reliance}
\end{figure*}

\subsection{Reliance Behaviour}
\edit{Following the methods of previous work \cite{schemmer2023appropriate}}, we drop the data instances where the participant's initial answer agreed with the LLM advice for the reliance analysis, as it is not possible to measure deliberate reliance in those cases. \edit{For Numerical Estimation, we do not use a strict definition of agreement, as it is unlikely for a participant’s initial estimate to exactly match the LLM advice. Instead, we adopt a heuristic of \textit{perceived agreement}. For example, a participant may perceive a guess of 700 to be ``pretty much the same" as an advice of 750. We use a conservative 10\% boundary for perceived agreement, as participants were primed with the expectation that guesses within $\pm$10\% of the true count would earn a bonus, making this range a reasonable proxy for perceived agreement.\footnote{\edit{We refine this heuristic with caveat that for smaller levels of advice, the boundary of perceived equivalence becomes very small. To address this, we include an additional $\pm$ 50 padding around the advice. As a robustness check, we also conduct reliance analysis using a 5\% agreement boundary without additional padding and find similar significance trends to those reported.}}} After excluding the instances of agreement, 87.1\% of samples in the LSAT task and 88.8\% in Numerical Estimation are retained. 

To answer \textbf{RQ1}, we compute the Relative LLM Reliance (RLR), Relative Self-Reliance (RSR), and Appropriate Reliance Ratio (ARR) across all data instances collected per task in each condition. For interpretation, RLR indicates the rate of reliance on the LLM when the advice is beneficial, and RSR indicates the rate of self-reliance when the LLM advice is bad. ARR is the ratio of RLR to ($1-$RSR), where higher values signify more appropriate reliance on the LLM (less over-reliance and less under-reliance, or at least a positive trade-off such as much less over-reliance and only a small increase in under-reliance). In terms of reliance outcomes, over-reliance maps to low RSR (i.e.\ relying on the LLM when the advice is bad), under-reliance maps to low RLR (i.e.\ not relying on the LLM when the advice is good), and appropriate reliance maps to when both are high. Figure \ref{fig:reliance_reference}, reconstructed based on Figure 1 from \cite{schemmer2022should}, provides a visual reference for the relationship between reliance outcomes and RLR and RSR values. Table \ref{tab:reliance} displays all the metrics calculated for each condition, separately for the LSAT and Numerical Estimation tasks. For the interventions, there are also \textit{p}-values indicating the significance of the metric difference between treatment and \control.

\begin{table*}[]
\caption{Summary of reliance metrics (RLR, RSR, and ARR) calculated for all four conditions in both LSAT and Numerical Estimation tasks, where the latter has results expressed at three different levels of WoA thresholds. Significance is calculated using permutation testing between \control \space and each intervention, where significant differences are \textbf{bolded}. Quantities with a positive change are further indicated with \textbf{(+)}. }
\label{tab:reliance}
\centering
\begin{tabular}{cccccc}
\toprule
& & \textbf{Control} & \textbf{Reliance Disclaimer} & \textbf{Uncertainty Highlighting} & \textbf{Implicit Answer} \\ 
\midrule

\multicolumn{1}{c}{\multirow{3}{*}{\textbf{LSAT}}}   
& \textit{RLR} & 0.66 & 0.71, p=.2 & 0.52, p=.1 & 0.58, p=.2 \\ 
& \textit{RSR} & 0.44 & 0.56, p=.05 & \textbf{0.62, p=.01 (+)} & \textbf{0.63, p=.002 (+)} \\ 
& \textit{ARR} & 1.17 & \textbf{1.62, p=.04 (+)} & 1.35, p=.2 & 1.54, p=.09 \\ 
\midrule

\multicolumn{1}{c}{\multirow{3}{*}{\begin{tabular}[c]{@{}c@{}}\textbf{Numerical}\\ \textbf{Estimation}\\ (WoA \textgreater 0.01)\end{tabular}}}
& \textit{RLR} & 0.91 & 0.91, p=.4 & 0.93, p=.3 & \textbf{0.67, p\textless{}.001 (–)} \\ 
& \textit{RSR} & 0.15 & 0.22, p=.1 & 0.23, p=.1 & \textbf{0.43, p=.004 (+)} \\ 
& \textit{ARR} & 1.07 & 1.17, p=.2 & 1.20, p=.1 & 1.18, r=.2 \\ 
\midrule

\multicolumn{1}{c}{\multirow{3}{*}{\begin{tabular}[c]{@{}c@{}}\textbf{Numerical}\\ \textbf{Estimation}\\ (WoA \textgreater 0.5)\end{tabular}}}
& \textit{RLR} & 0.8 & 0.74, p=.2 & 0.8, p=.5 & 0.45, p=.17 \\ 
& \textit{RSR} & 0.39 & 0.47, p=.1 & 0.42, p=.3 & \textbf{0.55, p=.007 (+)} \\ 
& \textit{ARR} & 1.32 & 1.38, p=.4 & 1.37, p=.4 & 1.01, p=.1 \\ 
\midrule

\multicolumn{1}{c}{\multirow{3}{*}{\begin{tabular}[c]{@{}c@{}}\textbf{Numerical}\\ \textbf{Estimation}\\ (WoA \textgreater 0.99)\end{tabular}}}
& \textit{RLR} & 0.47 & 0.39, p=.2 & \textbf{0.3, p=.02 (–)} & 0.34, p=.05 \\ 
& \textit{RSR} & 0.68 & \textbf{0.79, p=.02 (+)} & 0.77, p=.05 & 0.67, p=.5 \\ 
& \textit{ARR} & 1.45 & 1.91, p=.2 & 1.34, p=.4 & 1.03, p=.1 \\ 
\bottomrule
\end{tabular}
\end{table*}

\xhdr{LSAT}
The visual representations of the metrics and their significances for LSAT are plotted in Figure \ref{fig:reliance_mc}, where \control\space is plotted in gray and we compare the interventions against it. We find that \discl \space is the \textit{only intervention} that has a significantly positive influence on the ARR, by reducing both over-reliance and under-reliance (i.e.\ it is higher and further to the right than \control). \hl \space and \implicit \space both significantly improve self-reliance, but at the cost of shifting reliance away from the LLM when following it would be appropriate (although this decrease is not statistically significant)---these interventions tend to reduce LLM reliance overall, both when the advice is bad and when it is good. Both interventions have higher ARR values than \control, but the difference is not significant at our power level. These results indicate that a \textbf{\textit{static, persistent disclaimer message}} can induce appropriate reliance more effectively than more specialized methods.

\xhdr{Numerical Estimation}
This task requires more care to analyze, as WoA is a continuous rather than binary measure of reliance. As such, we compute reliance behaviours at several different WoA thresholds:
\begin{itemize}[noitemsep,topsep=0pt]
    \item \textit{WoA > 0.01}: Any adjustment towards the LLM's advice is counted as reliance (highest sensitivity). Under this definition, 79.7\% of trials count as reliant behaviour.
    \item \textit{WoA > 0.5}: Adjustments must be made over 50\% towards the LLM's advice to count as reliance (moderate sensitivity). Under this definition, 60.7\%  of trials count as reliant behaviour.
    \item \textit{WoA > 0.99}: Adjustment towards the LLM's advice must be nearly total to count as reliance (lowest sensitivity). Under this definition, 32.7\% of trials count as reliant behaviour.
\end{itemize}

Values and significance for RLR, RSS, and ARR are expressed at these three WoA threshold levels in both Table \ref{tab:reliance} and Figure \ref{fig:reliance_ne}. Unlike in the LSAT task, we do not observe any intervention to effectively improve appropriate reliance. Across different WoA thresholds, \implicit \space increased appropriate self-reliance the most (significant at WoA thresholds of 0.01 and 0.5). At the lowest WoA threshold, this also increases under-reliance---which is undesirable. This could be due to being forced to perform simple math calculations to see the advised number---which people may not do if they are satisfied with their own answer, or believe that extracting the advice requires more effort than the reward is worth.
\hl \space also causes significant decrease in RLR at high WoA thresholds, indicating that people are hesitant to fully use the advice.
This corroborates Figure \ref{fig:NE_woa}, which shows that \hl \space and \discl \space both have reduced density of WoA values close to 1 compared to \control. 

Another key difference between the Numerical Estimation and LSAT tasks is the length and detail of the advice. A higher barrier to cognitive engagement may lead to hetereogenous outcomes in how participants approach the task. For instance, some participants might adopt the advice as a shortcut to make easy, satisficing decisions, while others might disregard the advice due to its excessive length and complexity \cite{eigner2024determinants}. The discrepancy in results between our two tasks also warns that results of interventions \textit{may not generalize} beyond the specific tasks they are tested on, emphasizing the need for thorough validation across diverse tasks before broader deployment.

\begin{figure*}[t!]
    \centering
    \begin{subfigure}[t]{0.4\linewidth}
        \centering
        \includegraphics[width=\linewidth]{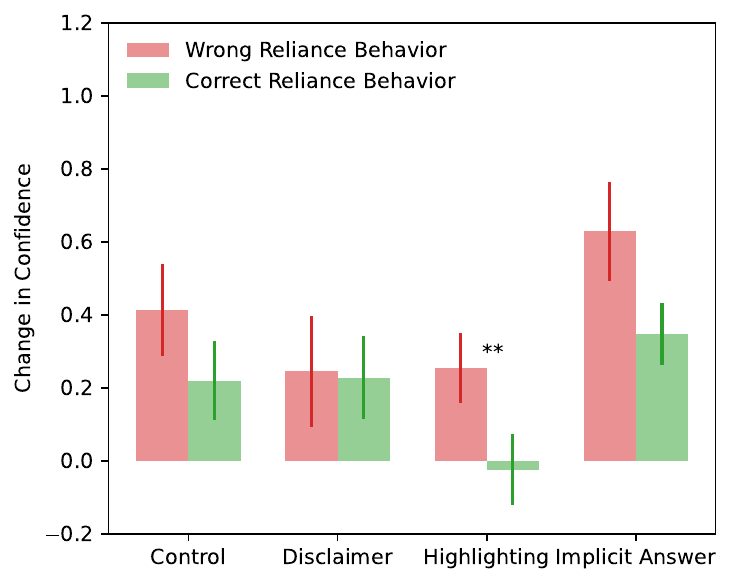}
        \caption{Correct (PL, PS) vs. Wrong (NL, NS) Reliance.}
        \label{fig:MC_reliance_conf}
    \end{subfigure}%
    ~ 
    \hspace{2em}
    \begin{subfigure}[t]{0.4\linewidth}
        \centering
        \includegraphics[width=\linewidth]{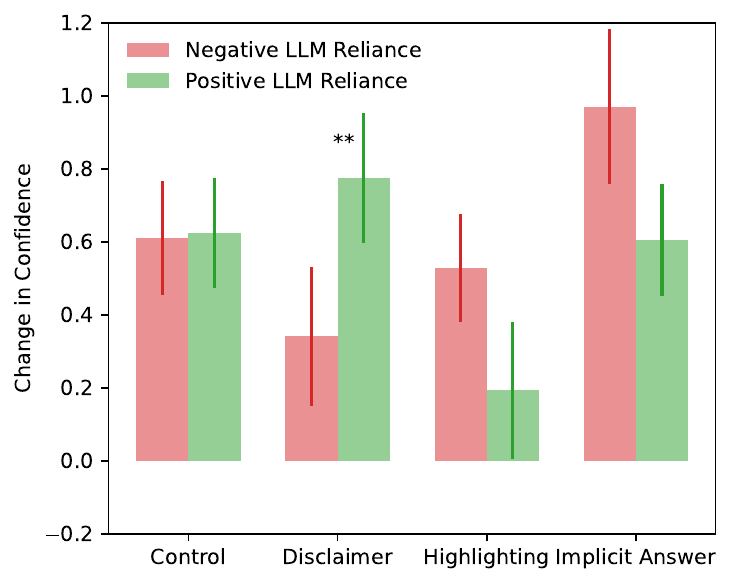}
        \caption{Positive LLM Reliance vs. Negative LLM Reliance.}
        \label{fig:MC_TP_FP_conf}
    \end{subfigure}
    \caption{\edit{Change in confidence at the question level for LSAT}.}
    \label{fig:MC_conf}
\end{figure*}

\begin{figure*}[t!]
    \centering
    \begin{subfigure}[t]{0.4\linewidth}
        \centering
        \includegraphics[width=\linewidth]{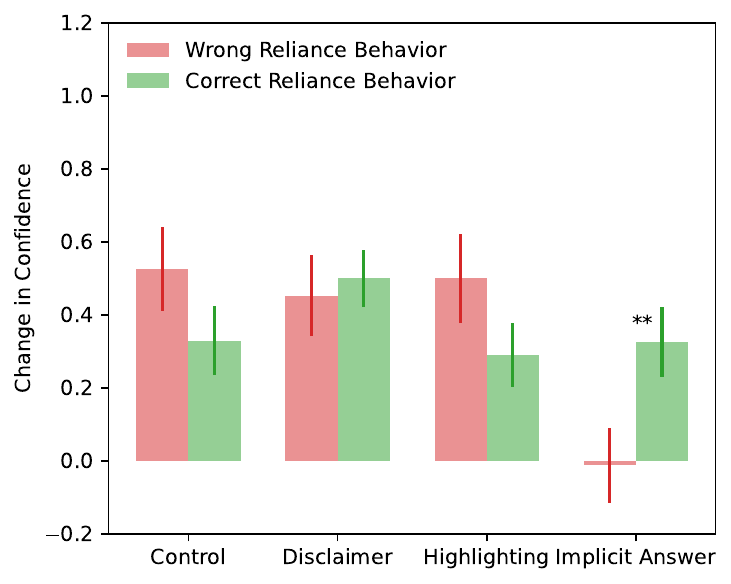}
        \caption{Correct (PL, PS) vs. Wrong (NL, NS) Reliance.}
        \label{fig:NE_reliance_conf_0.5}
    \end{subfigure}%
    ~ 
    \hspace{2em}
    \begin{subfigure}[t]{0.4\linewidth}
        \centering
        \includegraphics[width=\linewidth]{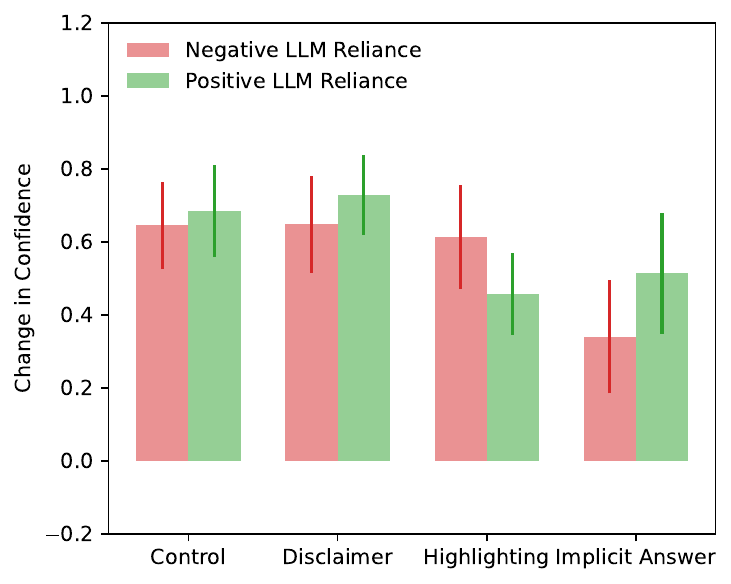}
        \caption{Positive LLM Reliance vs. Negative LLM Reliance.}
        \label{fig:NE_TP_FP_conf_0.5}
    \end{subfigure}
    \caption{Change in confidence at the question level for Numerical Estimation with a WoA threshold of 0.5.}
    \label{fig:NE_conf}
\end{figure*}

\subsection{Confidence Calibration and Perceptions}
\begin{figure}
    \centering
    \includegraphics[width=0.8\linewidth]{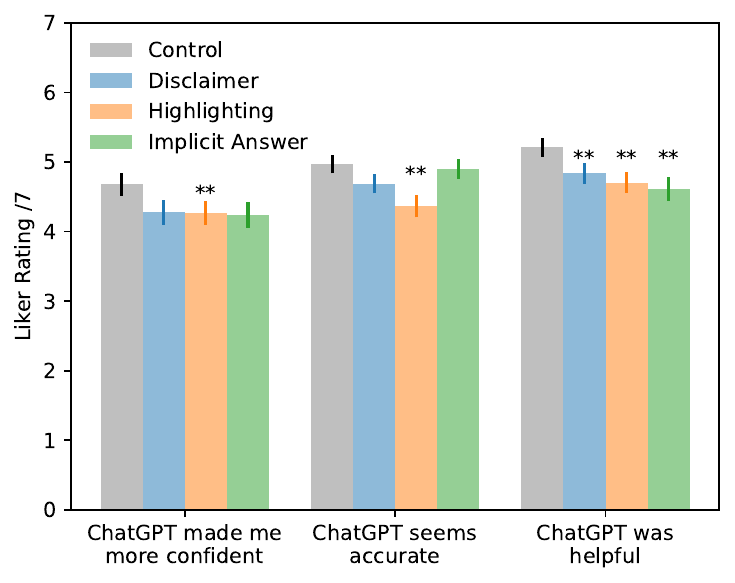}
    \vspace{-1em}
    \caption{Post-experiment perceptions ratings. \edit{Overall, subjective perceptions are lower for the intervention conditions than Control, possibly due to added usability friction. (**) indicates interventions that scored significantly lower than Control in each respective perception category}.}
    \label{fig:perceptions}
\end{figure}

\begin{figure*}[t!]
    \centering
    \begin{subfigure}[t]{0.4\linewidth}
        \centering
        \includegraphics[width=\linewidth]{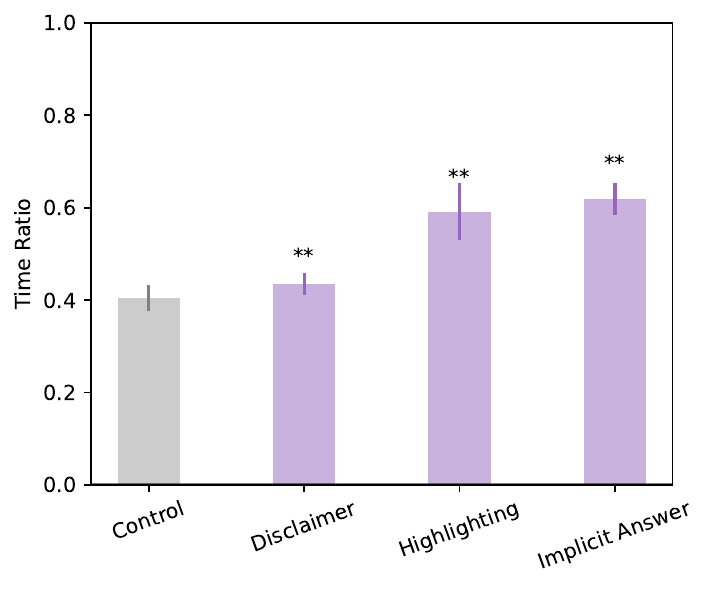}
        \caption{LSAT time results.}
        \label{fig:MC_time}
    \end{subfigure}%
    ~ 
    \hspace{2em}
    \begin{subfigure}[t]{0.4\linewidth}
        \centering
        \includegraphics[width=\linewidth]{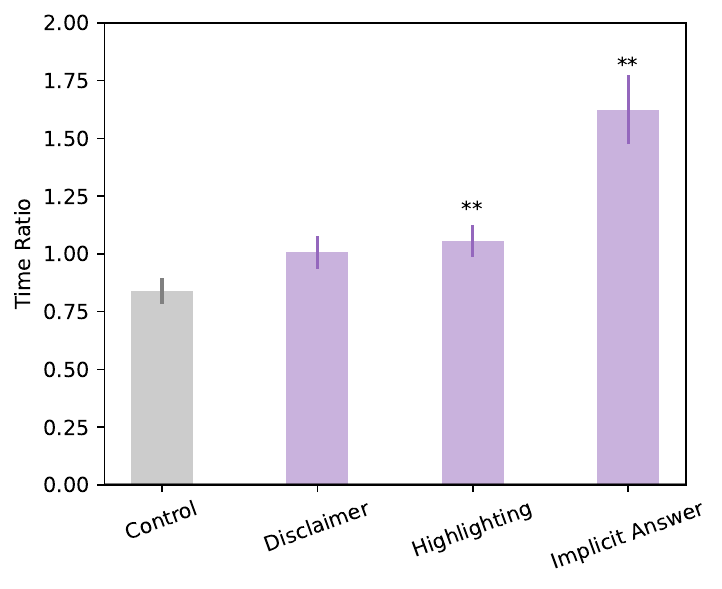}
        \caption{Numerical Estimation time results.}
        \label{fig:NE_time}
    \end{subfigure}
    \caption{Relative time spent at the LLM advice stage compared to the initial question stage. Note that the scale of the y-axes are different between tasks.}
    \label{fig:time}
\end{figure*}

 To address \textbf{RQ2}, we examine participant confidence in \edit{their pre-advice answer and their post-advice answer} for each question, along with several post-experiment Likert scale ratings of their subjective perceptions. We compare the changes in confidence between cases where participants made correct reliance decisions with those where they made incorrect ones, and further focus on instances of positive versus negative LLM reliance (over-reliance). Unless otherwise indicated, significance is calculated \textit{within} each condition using the Mann-Whitney U test with $p<.05$ as the significance threshold. \edit{Error bars in the bar plots of Figures \ref{fig:MC_conf}-\ref{fig:time} represent standard errors.}

\xhdr{LSAT}
Figure \ref{fig:MC_reliance_conf} shows participants' LSAT question confidence changes for correct and wrong reliance behaviours. The most salient takeaway is that on-aggregate across all conditions (not explicitly shown in the plot), \textit{people get more confident when they over- and under-rely} than when making the appropriate reliance decisions ($p=.02$). Within each condition, \hl \space is the only one where this observed effect is significant ($p=.03$), which indicates poor confidence calibration since it is undesirable for people to be more confident in wrong decisions.

As an extension, Figure \ref{fig:MC_TP_FP_conf} specifically focuses on cases where people chose to rely on the LLM. Here, the confidence scores tend to increase more as people trusted the LLMs enough to use them. We observe that \discl \space is the only condition that achieves \textit{calibrated confidence}---the increase in confidence from using the LLM's advice is significantly lower when the participant over-relied than when they relied appropriately ($p=.04$). Though not significant, it is interesting to observe that the opposite trend can be demonstrated in the other interventions, where confidence increases more in cases of over-reliance. This indicates that reliance interventions that operate by moderating users' confidence in the LLM may inadvertently \textit{miscalibrate} confidence with respect to appropriate reliance.

\xhdr{Numerical Estimation}
For this task, we repeat the confidence analyses and show them in Figure \ref{fig:NE_conf}. For simplicity, reliance is determined based on the middle WoA threshold of 0.5, since the confidence outcomes between high and low WoA thresholds are qualitatively similar (the results for WoA thresholds of 0.01 and 0.99 are shown in Appendix \ref{app:woa_conf} as Figures \ref{fig:NE_conf_0.01}-\ref{fig:NE_conf_0.99}). 
\implicit \space is the only condition to achieve a significant difference in calibrating confidence changes to be higher in correct reliance outcomes. However, no significant differences are seen in the behaviours when we select for instances that explicitly result in reliance on LLMs, which may indicate that \implicit \space is effective at reducing confidence for \textit{negative self-reliance}---perhaps people recognize that they can benefit from the advice, but did not want to go through the effort of calculating the recommendation.

\xhdr{Overall Post-Perception}
Lastly, Figure \ref{fig:perceptions} shows the mean Likert ratings per perceptions category by condition. The full subjective perception statements are shown in the figure's labels (1=disagree, 7=agree). We calculate significance between the \control \space and the interventions using the Mann-Whitney U test with $p<.05$ as threshold. Although all LLM conditions scored positively across the three categories, \control \space scored highest in subjective perceptions compared to all other the interventions. In particular, \hl \space is the only intervention that had significantly lower ratings in all categories compared to \control, indicating that our implementation may have induced too much uncertainty in the users. An alternative approach to moderate this effect may be to highlight highly certain tokens in green in addition to uncertain tokens in red. \edit{In addition, we find that the frequency of LLM usage correlates positively with all three subjective perception measures, but not with reliance behaviours -- see Appendix \ref{app:usage} for details}. The main implication here is that uncertainty calibration methods can have the undesirable outcome of inducing negative perceptions of accuracy and usefulness of the LLM, which increases overall algorithm aversion against LLM advice, rather than appropriate reliance.

\subsection{Time of Engagement}
We compare the relative time of engagement with given LLM advice by taking the ratio of time spent in the second stage to the first stage. Time is correlated to cognitive engagement, and previous studies have found that introducing slight temporal frictions like waiting for delayed AI advice has potential to affect user behaviour \cite{buccinca2021trust, collins2024modulating}. Figure~\ref{fig:time} shows the time ratio by condition for both the LSAT (Figure \ref{fig:MC_time}) and Numerical Estimation (Figure \ref{fig:NE_time}) tasks, where significance is indicated by Mann-Whitney U test with $p<.05$. Overall, participants spent more time in the Numerical Estimation question, with the average relative time ratio exceeding 1, potentially due to the average length of the advice being longer. On the other hand, the time ratio is well below 1 in LSAT.

In both tasks, the order of relative time spent on the advice roughly matches \edit{the effort required to extract the answer from the advice}. \discl \space is closest to matching the time ratio of \control \space while \implicit \space takes much longer, reaching nearly double the relative time in Numerical Estimation. \hl \space is somewhere in between, but also with significant increase in time over \control. Taken with the results on the efficacy of \discl, our results indicate that longer engagement time does not necessarily correlate with reliance outcomes. The easier, more digestible intervention had the most tangible effect on appropriate reliance. 
Perhaps with more training and rigour, especially for \hl, participants can become more proficient at interpreting advice with more complex uncertainty calibration.

\section{Discussion}
\label{sec:discussion}

\edit{We now present the key findings to our research questions on appropriate reliance and confidence calibration. We then extend our findings to broader implications for design and for evaluating reliance on LLMs, and discuss the limitations of our experiments. }

\subsection{\edit{Key Findings}}

\xhdr{\edit{Cognitive Forcing Designs Reduce Over-Reliance, but May Not Improve Appropriate Reliance}}
In our study, we investigated the effectiveness of three interventions on LLM reliance in two diverse and challenging tasks. \edit{We use measures of inappropriate reliance, separating over-relying on bad advice from under-relying on helpful advice, to uncover mechanisms of sub-optimal human-LLM team performance. }
Our findings suggest that cognitive forcing designs in the form of confidence calibration (highlighting uncertain tokens) and usability friction (adding disclaimers and removing direct answers) \textit{can} help people shift away instinctively relying on the LLM, but improving appropriate advice usage is a harder challenge.

\edit{While the effect of each intervention varied by the task, \textit{all three achieved a reduction in over-reliance} in at least some experiment configurations. However, the only intervention to  reduce \textit{both over-reliance and under-reliance} is \discl \space in the LSAT logical reasoning task.}
\edit{The Numerical Estimation task captured reliance with more granularity, which provided insight on how the interventions induced under-reliance differently. Our results suggest that the added friction of calculating the answer in \implicit \space caused people to disproportionately discard the LLM advice (not rely at all), while the confidence calibration in \hl \space caused people to hesitate to fully follow advice (rely, but not fully). 
}

\edit{The increase in advice under-reliance supports the cost-benefit framework from explainable AI literature, which shows that users will not engage with explanations if they are more cognitively complex than the task itself \cite{vasconcelos2023explanations}. 
This suggests that for the LSAT task, static \discl s may have added just enough friction to dissuade habitual over-reliance.}
\edit{However, advice with other interventions, like \implicit \space in Numerical Estimation, may have been perceived as requiring more effort to use than it was worth. Future work should aim to quantify the comprehension difficulty of the interventions and the tasks to understand how they relate to reliance behaviours. In implementing interventions, designers should consider the targeted use case and develop cognitive forcing techniques that do not increase burden substantially.
}

\xhdr{\edit{Confidence When Using LLMs is Poorly Calibrated}} \edit{We examine how participants' confidence in their answers change after seeing the LLM's advice. 
Ideally, even with incorrect reliance behaviour, confidence should be lower when individuals are internally uncertain about their decision, such as when they are skeptical of the LLM’s advice. However, we largely observe the opposite or inconclusive effects, where wrong reliance results in higher gains in confidence. This raises particularly problematic implications, suggesting that individuals were \textit{overall uncalibrated when receiving assistance from LLMs}, corroborating prior findings \cite{zhang2023taking, si2023large}}

\edit{There were a few areas where the interventions had success. 
In the Numerical Estimation task, \implicit \space reduced confidence during under-reliance, while in the LSAT task, \discl \space lowered confidence gains during over-reliance, which can be exploited as feedback mechanisms for users to learn from.
Both methods present different pathways using which confidence can be influenced by modifying LLM output, but further work is needed to achieve generalizable findings. As shown in prior works in human-AI collaboration, self-confidence and advice agreement rates have nuanced relationships with reliance on advice \cite{lu2021human}, which can be further investigated in the context of human-LLM collaboration. 
}

\edit{We also find perceptions of helpfulness were negatively impacted by all three interventions, while perceived self-confidence and advice accuracy were only reduced by \hl. Confidence calibration interventions may risk increasing a user's aversion towards LLMs rather than encouraging appropriate evaluation if overall performance is perceived as unacceptably low, or if the intervention makes the advice difficult to understand and use. 
Further work can explore qualitative factors of what participants disliked or found more cumbersome, and how their perception of their internal decision-making process was affected. 
}

\subsection{\edit{Broader Implications}}
\edit{With the integration of Copilot in consumer PCs, the widespread use of LLMs in customer service and information retrieval, and advances in Generative AI for creative work, interactions between end users and these tools are becoming nearly unavoidable. 
As Generative AI advances rapidly, consumers' literacy in its appropriate use is likely to lag behind, raising concerns about how to ensure proper use of the many AI-powered tools and products being introduced \cite{annapureddy2024generative}.
In our study, we show that wrongful reliance is frequent and users are not calibrated to be aware of their behaviour. Interventions that reduce over-reliance can be have undesired effects of increasing under-reliance, and require thorough validation in human user studies. We highlight the need to extend the appropriate reliance metrics to realistic usage scenarios.}

\xhdr{\edit{Generalization of the Appropriate Reliance Metrics}}
\edit{Through our experiments, we highlight the flexibility of the Relative LLM Reliance, Relative Self-Reliance, and Appropriate Reliance Ratio metrics in benchmarking interventions across two diverse task formats. 
While prior works have focused on binary or multiple choice answers, we demonstrate the suitability of the metrics for numerical targets, which opens up their applicability to additional real-world tasks. 
These measures enable more nuanced analyses of suboptimal human-LLM collaboration beyond task performance alone, and we recommend their incorporation into future empirical research on LLM usage.
The metrics can be generalized to any LLM-assisted task where there is a deterministic notion of optimality (i.e. some outcomes are preferred over others), such as code compilation success for programming, accuracy of arguments in writing persuasive essays, and requirement satisfaction in travel planning. Extending experiments to more open-ended tasks requires further nuance to faithfully capture agreement, reliance, and correctness. However, we theorize some degree of generalization in our high level findings---that cognitive forcing interventions will cause people to approach reliance on LLM content more cautiously.}

\xhdr{LLM Design Trade-Offs Based on the Context of Use}
\edit{The question, \textit{``How can interventions be designed successfully?"}, still remains open. We uncover complex heterogeneity in how interventions performed across different tasks, indicating that optimal designs are likely context-dependent rather than universal. Our work motivates further investigation into the theoretical foundations linking individual factors, tasks characteristics, and LLMs' response styles, and towards developing a reliable empirical framework to evaluate novel interventions rigorously. }
Additionally, as our results suggest a trade-off between reducing over-reliance and increasing under-reliance, it is therefore vital that practitioners consider the \textit{context of use} of the LLM tool. This involves a closer evaluation of the stakeholders, users, and risks involved in the decision-making tasks \cite{liao2022connecting}. 

\edit{
In medical or legal decision-making, the heightened risk requires accountability, and errors from over-reliance may
cause extreme damage if unchecked \cite{cheong2024not, yang2023large}. 
In creative tasks where the diversity of humans is valued, over-reliance may introduce an undesirable degree of homogeneity \cite{kumar2024human}. Within the education space, over-reliance brings about another issue: the unintended reduction of skill development due to relying on the LLM as a cognitive crutch \cite{zhai2024effects}. 
As AI capabilities advance and eventually eclipse human capabilities, avoiding under-reliance may be preferred in lieu of aversion. However, scaling ability requires scaling oversight, representing further challenges in human-AI collaboration that require carefully calibrated trust and reliance \cite{bowman2022measuring}. 
}

\xhdr{Human Validation for Technical Development} 
Ensuring the success of human-AI interactions requires bi-directional design and validation from both AI and HCI researchers \cite{shen2024towards, vodrahalli2022uncalibrated}. 
Technical AI research has progressed significantly towards understanding LLMs' reasoning pathways, reducing hallucinations, and calibrating the confidence of outputs \cite{ayyamperumal2024current, feng2024don, lin2022teaching, zhang2023siren}. 
However, such works often rely on automating evaluations by using existing benchmarks or artificially simulating users, considering only theoretical design desiderata and bypassing human evaluation. 
These evaluation methods do not ascertain whether these methods will have the desired effects when deployed to real users. For example, we showed that highlighting log probabilities to calibrate confidence did not have the intended effect on humans, who may have found the added information too confusing to optimally leverage. 
Not only did reliance and performance outcomes lack improvement, we also found that participants' confidence changes were poorly calibrated, with a significant increase in post-task confidence when LLM advice is used \textit{inappropriately}. 
We therefore encourage the broader research community, particularly the technical AI community, to capture multi-dimensional measures from real users.

\subsection{Limitations}
Overall, our results would be further strengthened by future research that explores LLM reliance in more naturalistic and long-term settings, offering a clearer picture of how users assess and use model outputs in the real world. \edit{Incorporating qualitative measures can also provided more nuanced feedback on \textit{how} the participants regarded the advice and made choices.}

\xhdr{\edit{Engagement of Participants in Tasks}} The generalizability of our results is primarily limited by our design of a lab experiment with crowdworkers recruited from Prolific.  Since participants knew that they were participating in an experiment, they may have inferred that they were being evaluated on LLM reliance and become more cautious about their reliance behaviours. Although we tried to emulate the risks and benefits of real-world decision-making by rewarding participants for correctness, this may have encouraged participants to ``hedge"; to maximize their chances of earning \textit{some} bonus payment by choosing different answers at each stage. \edit{Our two-stage question answering set-up also emulates a specific type of decision-making where users are already engaged in the task. In fact, having users answer questions independently before advice can be considered a form of cognitive forcing even without additional interventions \cite{ashktorab2024emerging}. As LLM usage becomes more common, it is expected that many users will bypass this critical engagement and rely on LLMs by default, which can reflect in reliance outcomes differently}.

\xhdr{\edit{External Validity of the Tasks}}
Our chosen tasks do not perfectly translate to the usage contexts for which people consult LLMs in the real world. While multiple-choice questions provided clean correctness and reliance metrics, real-world tasks predominantly consist of open-ended questions and generative tasks (such as writing and coding). 
Knowledge-based questions were avoided in this study to ensure that the questions would be accessible to participants; however, real questions posed to LLMs frequently require synthesis of more complex ideas and knowledge bases, which may be more susceptible to factual hallucinations.

\xhdr{\edit{Realism of Interacting with the LLM}}
Due to the nature of the survey study, our participants were unable to interactively ask follow-up questions or use sanity checks to verify the model's outputs and prevent consequential errors. To ensure clean reliance metrics, they saw fixed advice and were asked not to consult external resources. 
Although we used direct outputs generated through tailored LLM prompts whenever possible, we did apply manual adjustments in some interventions and artificially manipulated the recommendation in the LLM's advice. The distribution of advice quality was held constant regardless of the LLM's true tendencies when responding to the question to prevent question difficulty from being a confound. \edit{We only explore one specific LLM, GPT-4o from OpenAI, so chatbots like Claude\footnote{https://claude.ai/}, Copilot\footnote{https://copilot.microsoft.com/}, Gemini\footnote{https://gemini.google.com/}, and other open-source models may reflect different conversational styles that can implicitly affect user behaviour. We also did not explore varying the system prompts to steer the LLM into embodying different personas. Furthermore, the framing of the LLM (e.g., ChatGPT vs. a generic chatbot) can prime user expectations of the quality and accuracy of the responses.}

\xhdr{\edit{Selection of Interventions}}
\edit{To operationalize our research questions, we made certain design choices in the implementation of the interventions. For example, we used a naive method of highlighting based on token generation probability, while prior works have identified that likelihood of editing is a better indicator \cite{vasconcelos2023generation}.} Our evaluation of the interventions also did not include any extensive training or longer-term follow-ups. As such, the impacts of each intervention in our studies are not necessarily representative of how they would impact reliance if optimally implemented in the real world.  
In particular, it is not clear if static disclaimers would persistently encourage appropriate reliance in deployment where users may experience alarm fatigue \cite{cvach2012monitor}, \edit{whereas dynamic disclaimers may have a longer impact \cite{bo2024disclosures}}. 
We explored a very small space of potential interventions, and encourage further human-centered evaluation of other uncertainty calibration techniques (e.g., confidence score \cite{virk2024enhancing}, semantic expressions of uncertainty \cite{kim2024m, lin2022teaching, steyvers2024calibration}, and contrastive explanations \cite{si2023large}) and setting appropriate expectations of the LLM (e.g., model report card \cite{yang2024report} and usage tutorial \cite{chiang2022exploring}).

\section{Conclusion}
\label{sec:conclusion}
\edit{We measured the effects of three different LLM interventions 
on users' reliance behaviours and confidence levels in two difficult LLM-assisted question-answering tasks. 
We find that adding a persistent disclaimer is the most effective at improving appropriate reliance (in the LSAT task) without significantly increasing the time needed to engage with the LLM advice, while both highlighting uncertain tokens and removing direct answers only reduced over-reliance and increased the engagement time. 
In addition, people's confidence levels were poorly calibrated, generally exhibiting higher confidence increases when they made the wrong reliance decision. 
Based on our results, we call for the following:  rigorous and human-centered evaluation of novel LLM reliance interventions; consideration for trade-offs between over- and under-reliance based on the context of use; and future research in naturalistic human-LLM collaboration tasks. 
}

\begin{acks}
We thanks Nicholas Carlini for helpful discussions during the early phase of the project. We further thank Lev McKinney, Lillio Mok, Harsh Kumar, Nakwon Rim, Paula Aoyagui, Zixin Zhao, and members of the UofT Computational Social Science Lab for feedback on the experimental methods and the draft of the manuscript.
\end{acks}

\bibliographystyle{ACM-Reference-Format}
\bibliography{references}

\pagebreak 
\renewcommand\thefigure{\thesection.\arabic{figure}} 
\setcounter{figure}{0}  

\appendix
\onecolumn
\section{Task Question Examples}
\label{app:questions} 
\begin{figure*}[h]
    \centering
    \begin{subfigure}[t]{.65\linewidth}
        \includegraphics[scale=0.55]{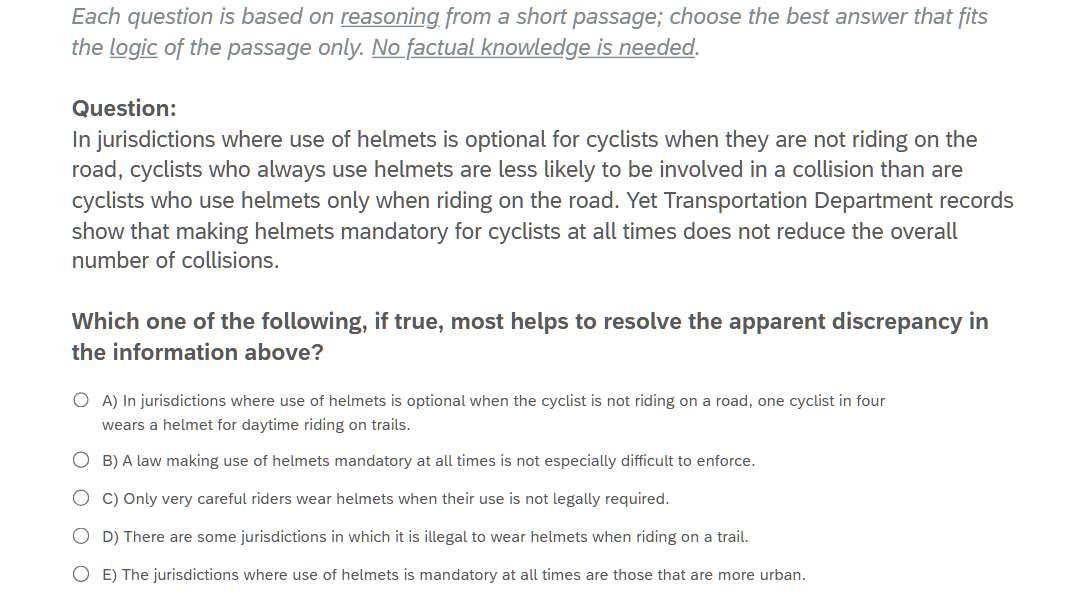}
        \caption{Stage 1 of the "Road" LSAT Logical Reasoning question.}
        \label{subfig:lr_q_road_stg1}
    \end{subfigure} 
    \begin{subfigure}[t]{0.65\linewidth}
        \vspace{0.5em}
        \includegraphics[scale=0.8]{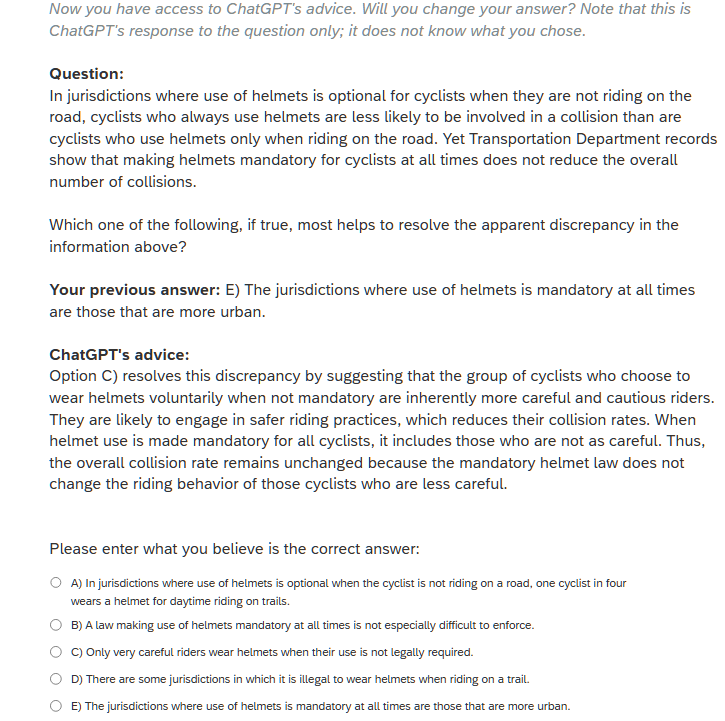}
        \caption{Stage 2 of the "Road" question with an example of base (\control) advice.}
        \label{subfig:lr_q_road_stg2}
    \end{subfigure} 
    
    \vspace{-1em}
    \caption{"Road" LSAT Logical Reasoning question}
    \label{fig:lr_q_road}
\end{figure*}

\begin{figure*}[h]
    \centering
    \begin{subfigure}[t]{.65\linewidth}
        \includegraphics[scale=0.8]{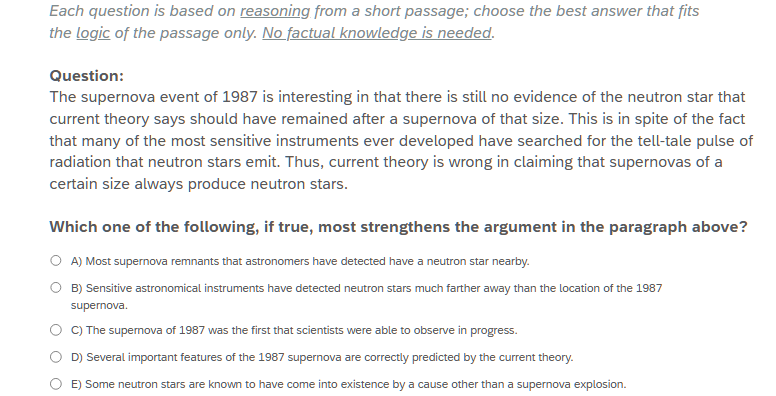}
        \caption{Stage 1 of the "Star" LSAT Logical Reasoning question.}
        \label{subfig:lr_q_star_stg1}
    \end{subfigure} 
    \begin{subfigure}[t]{0.65\linewidth}
        \vspace{0.5em}
        \includegraphics[scale=0.8]{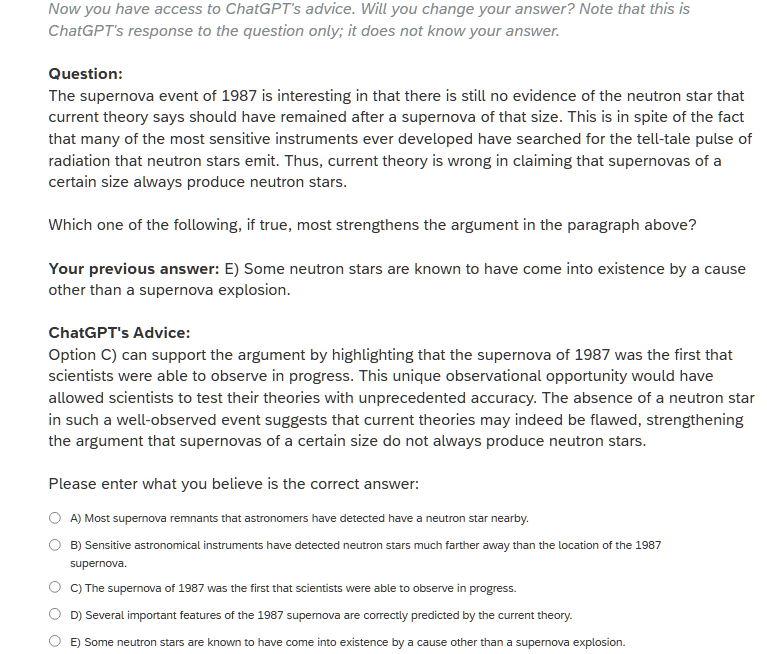}
        \caption{Stage 2 of the "Star" question with an example of base (\control) advice.}
        \label{subfig:lr_q_star_stg2}
    \end{subfigure} 
    
    \vspace{-1em}
    \caption{Screenshot of the "Star" LSAT Logical Reasoning question, taken from Qualtrics}
    \label{fig:lr_q_star}
\end{figure*}

Figures \ref{fig:lr_q_road} and \ref{fig:lr_q_star} show screenshots LSAT Logical Reasoning tasks questions. Participants answer independently in Stage 1, and then re-answer the same question with randomly sampled LLM advice in Stage 2. 

\begin{figure*}[h]
\centering
    \begin{subfigure}[t]{.8\linewidth}
        \includegraphics[scale=0.54]{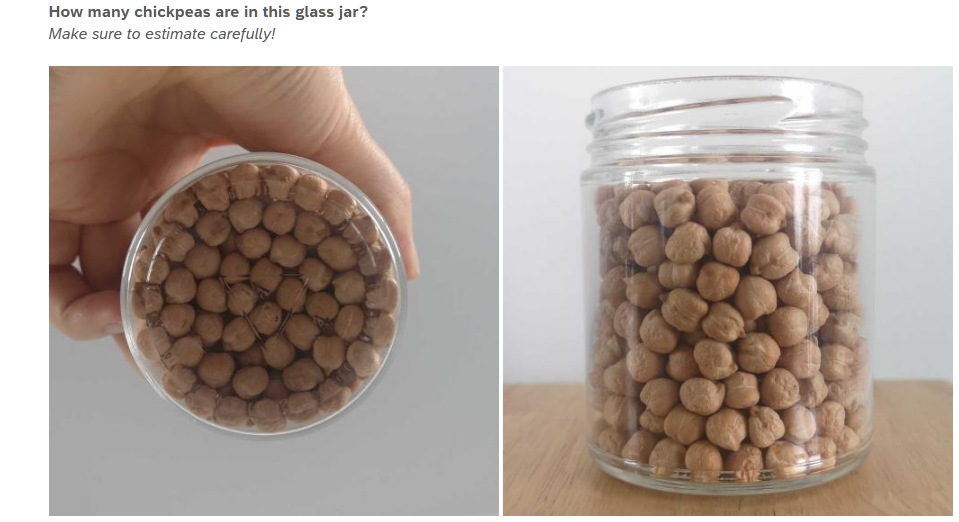}\\
        \includegraphics[scale=0.54]{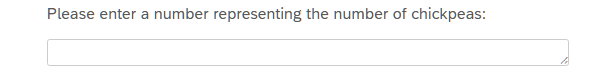} \\
       \vspace{-1em}
        \caption{Stage 1 of the chickpea Numerical Estimation question.}
        \label{subfig:ne_q_chickpea_stg1}
    \end{subfigure} 
    
    \begin{subfigure}[t]{0.8\linewidth}
            
       \vspace{1em}
        \includegraphics[scale=0.54]{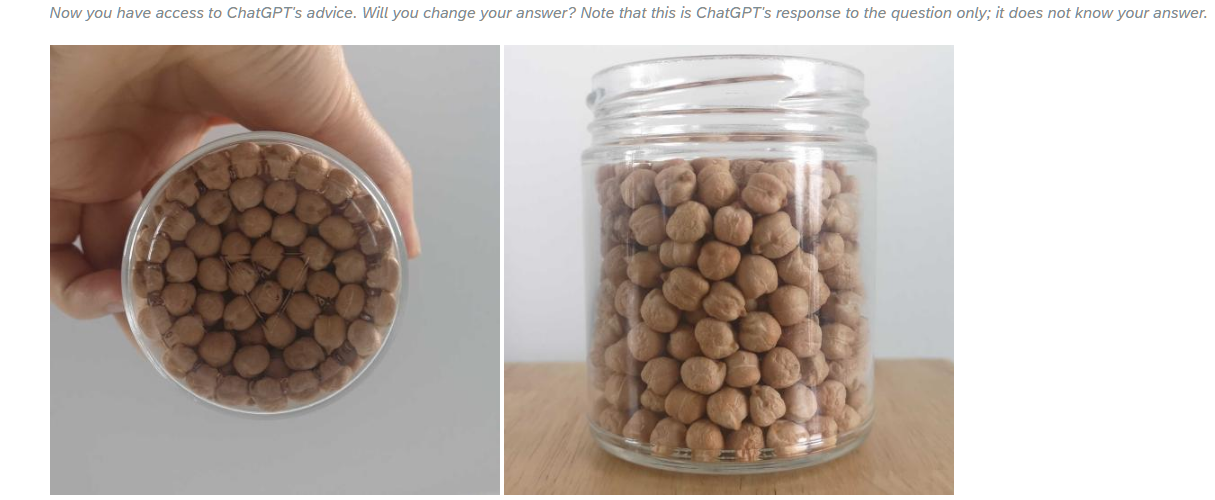}\\
        \includegraphics[scale=0.54]{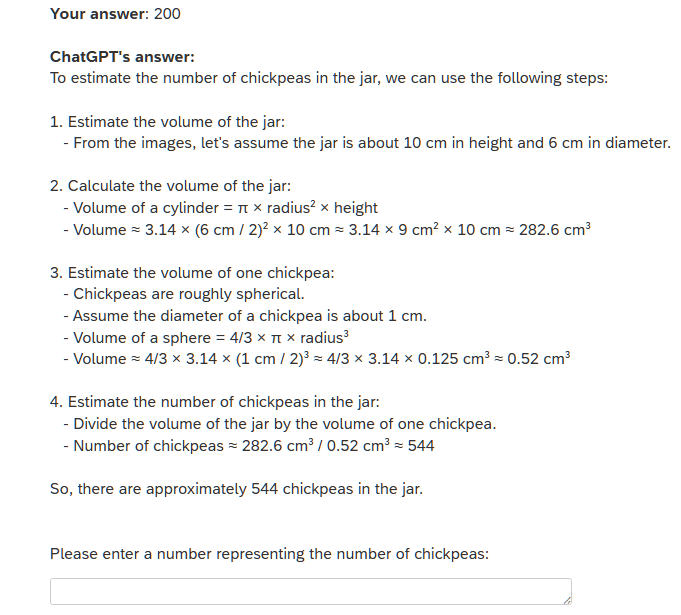} 
        \caption{Stage 2 of the chickpea question with an example of base (\control) advice.}
        \label{subfig:ne_q_chickpea_stg2}
    \end{subfigure} 
    \vspace{-1em}
    \caption{Screenshot of the chickpea Numerical Estimation question, taken from Qualtrics}
    \label{fig:ne_q_chickpea}
\end{figure*}

\begin{figure*}[t]
\centering
    \begin{subfigure}[t]{.8\linewidth}
        \includegraphics[scale=0.54]{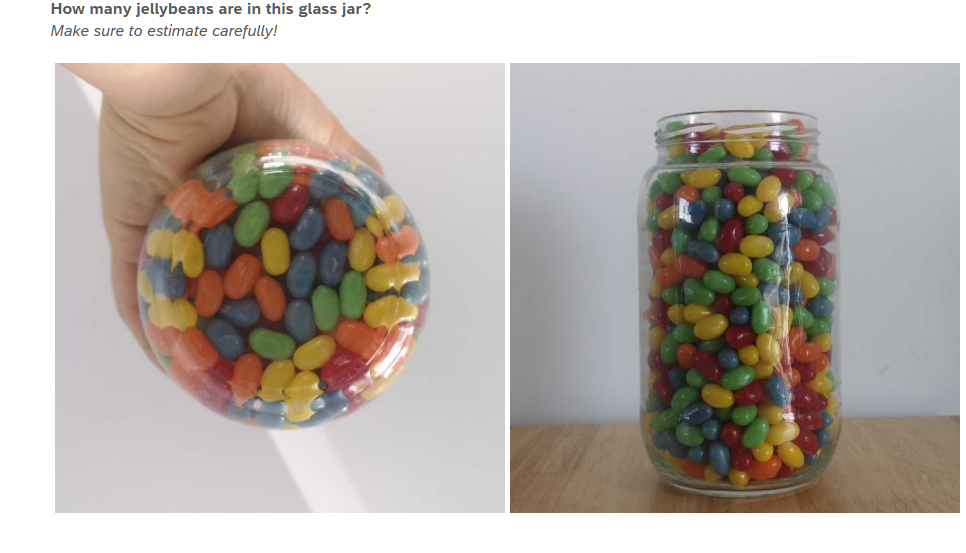}\\
        \includegraphics[scale=0.54]{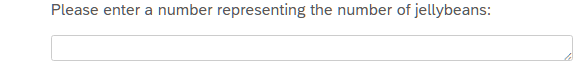} \\
       \vspace{-1em}
        \caption{Stage 1 of the jellybean Numerical Estimation question.}
        \label{subfig:ne_q_jellybean_stg1}
    \end{subfigure} 
    
    \begin{subfigure}[t]{0.8\linewidth}
            
       \vspace{1em}
        \includegraphics[scale=0.54]{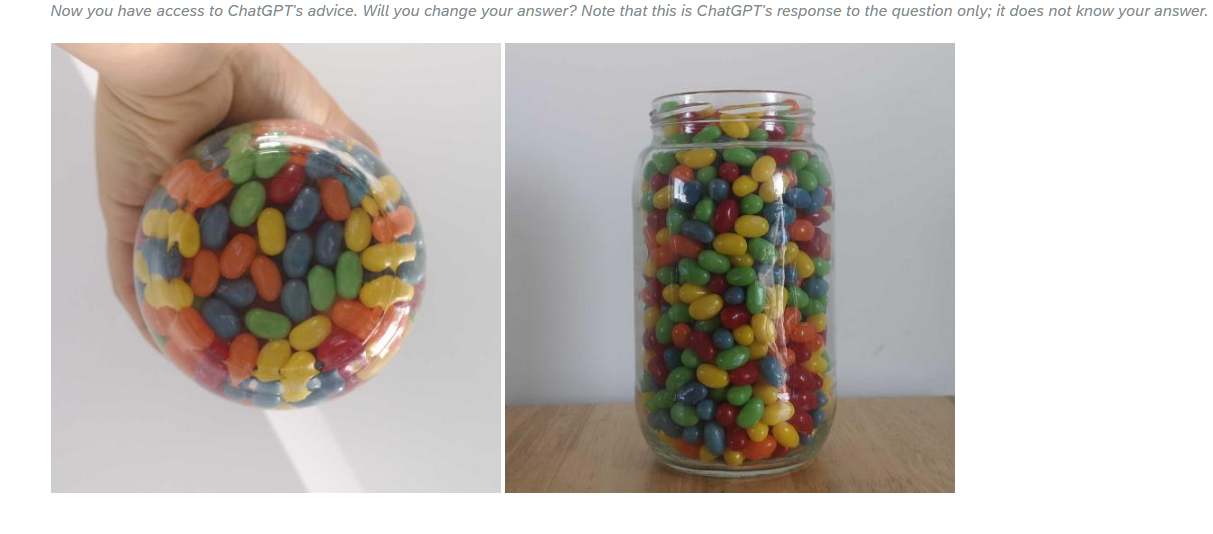}\\
        \includegraphics[scale=0.54]{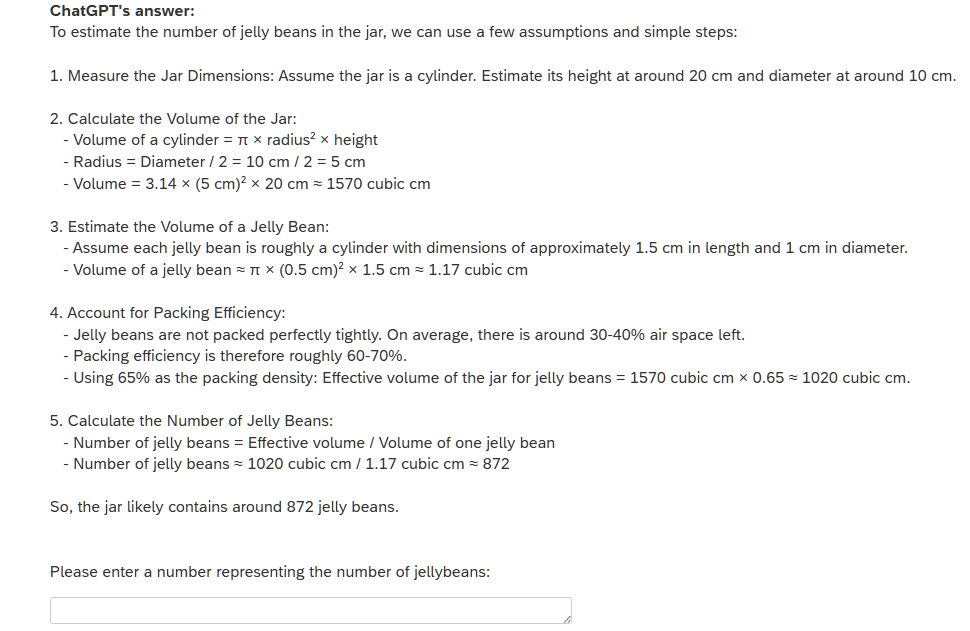} 
        \caption{Stage 2 of the jellybean question with an example of base (\control) advice.}
        \label{subfig:ne_q_jellybean_stg2}
    \end{subfigure} 
    \vspace{-1em}
    \caption{Screenshot of the jellybean Numerical Estimation question, taken from Qualtrics}
    \label{fig:ne_q_jellybean}
\end{figure*}

\pagebreak 
Figures \ref{fig:ne_q_chickpea} and \ref{fig:ne_q_jellybean} show screenshots of Numerical Estimation task questions, with the same two-stage design. 

\newpage

\clearpage

\pagebreak
\section{Control and Intervention Advice Examples}
\label{app:advice}
\begin{figure*}[h]
    \begin{subfigure}[t]{0.49\linewidth}
        \raggedright
        \includegraphics[scale=0.65]{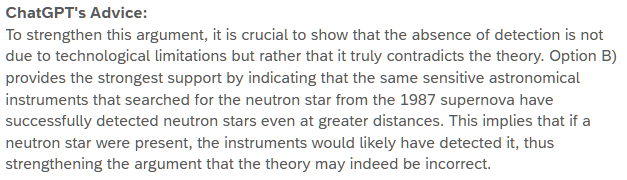}
        \label{subfig:lr_advice_ctrl}
        \caption{\control}
    \end{subfigure} 
    \begin{subfigure}[t]{0.49\linewidth}
        \raggedright
        \includegraphics[scale=0.65]{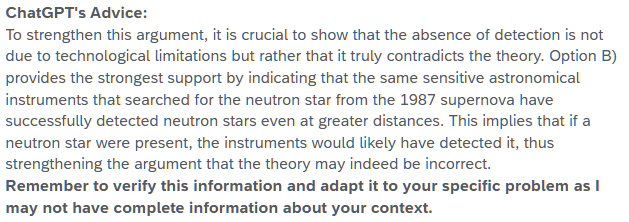}
        \label{subfig:lr_advice_disc}
        \caption{\discl}
    \end{subfigure} 
    \begin{subfigure}[t]{0.49\linewidth}
        \raggedright
        \includegraphics[scale=0.65]{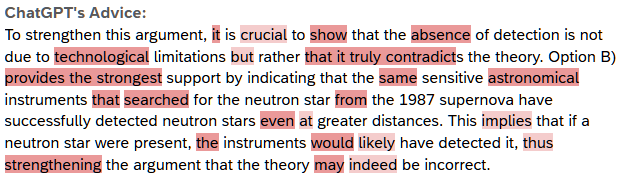}
        \label{subfig:lr_advice_hl}
        \caption{\hl}
    \end{subfigure}
    \begin{subfigure}[t]{0.49\linewidth}
        \raggedright
        \includegraphics[scale=0.65]{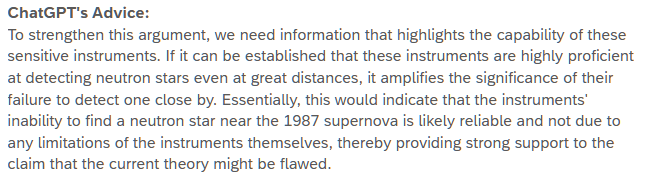}
        \label{subfig:lr_advice_expl}
        \caption{\implicit}
    \end{subfigure} 
    \caption{Sample LLM advice for all four conditions for an LSAT Logical Reasoning question.}
    \label{fig:lr_advice}
\end{figure*}

\begin{figure*}[h]
    \begin{subfigure}[t]{0.49\linewidth}
        \raggedright
        \includegraphics[scale=0.25]{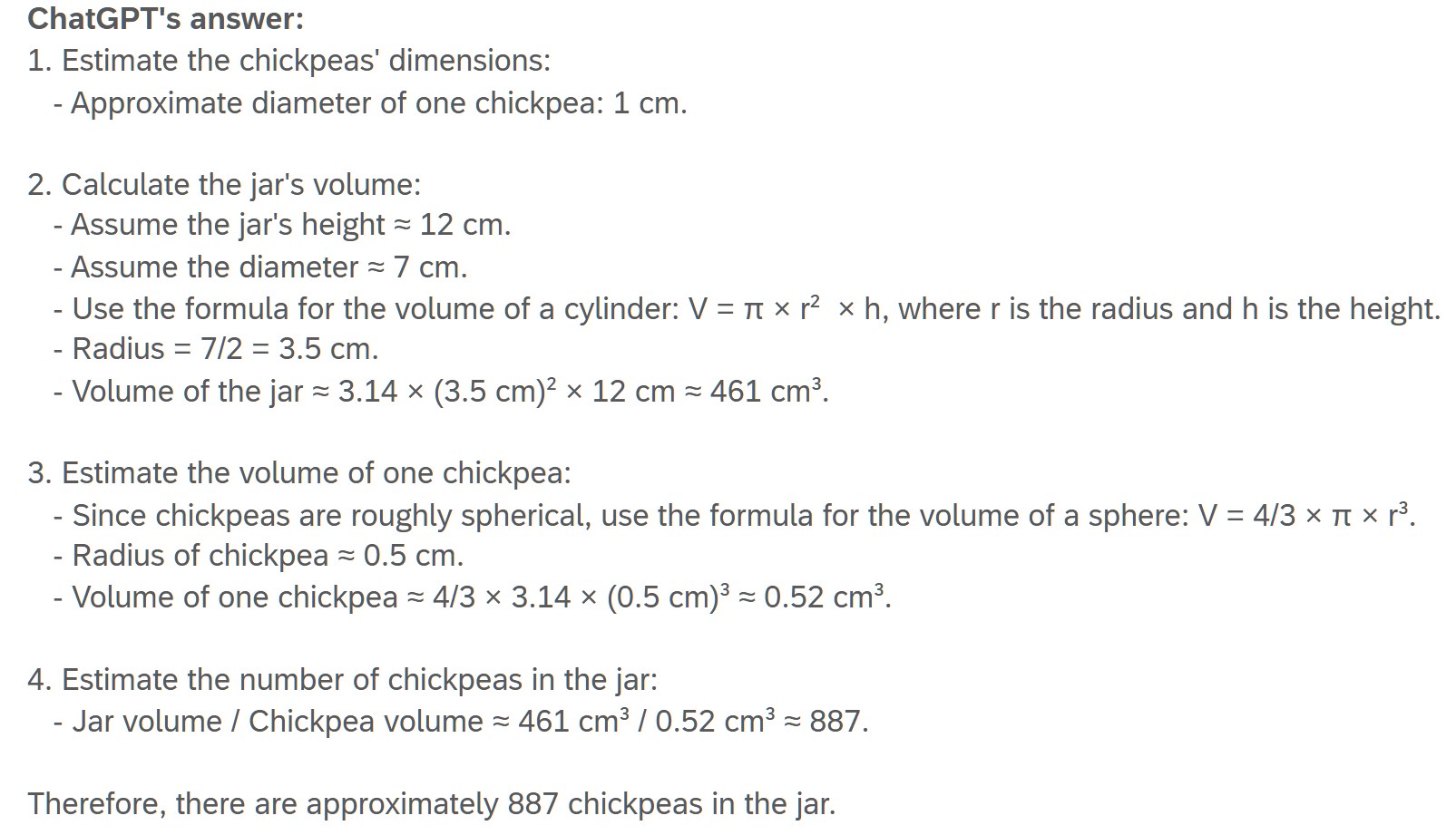}
        \label{subfig:ne_advice_ctrl}
        \caption{\control}
    \end{subfigure} 
    \begin{subfigure}[t]{0.49\linewidth}
        \raggedright
        \includegraphics[scale=0.25]{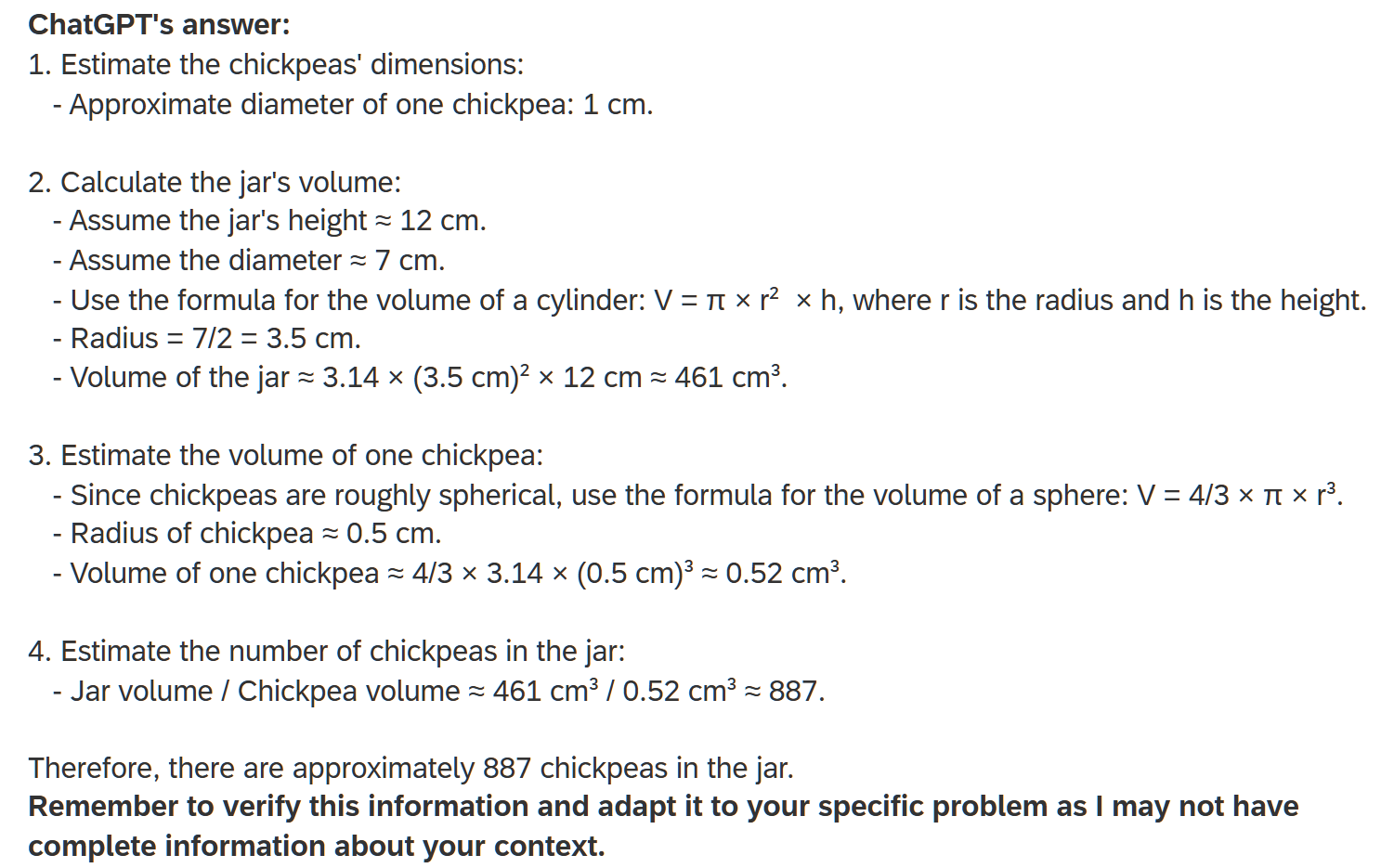}
        \label{subfig:ne_advice_disc}
        \caption{\discl}
    \end{subfigure} 
    \begin{subfigure}[t]{0.49\linewidth}
        \raggedright
        \includegraphics[scale=0.25]{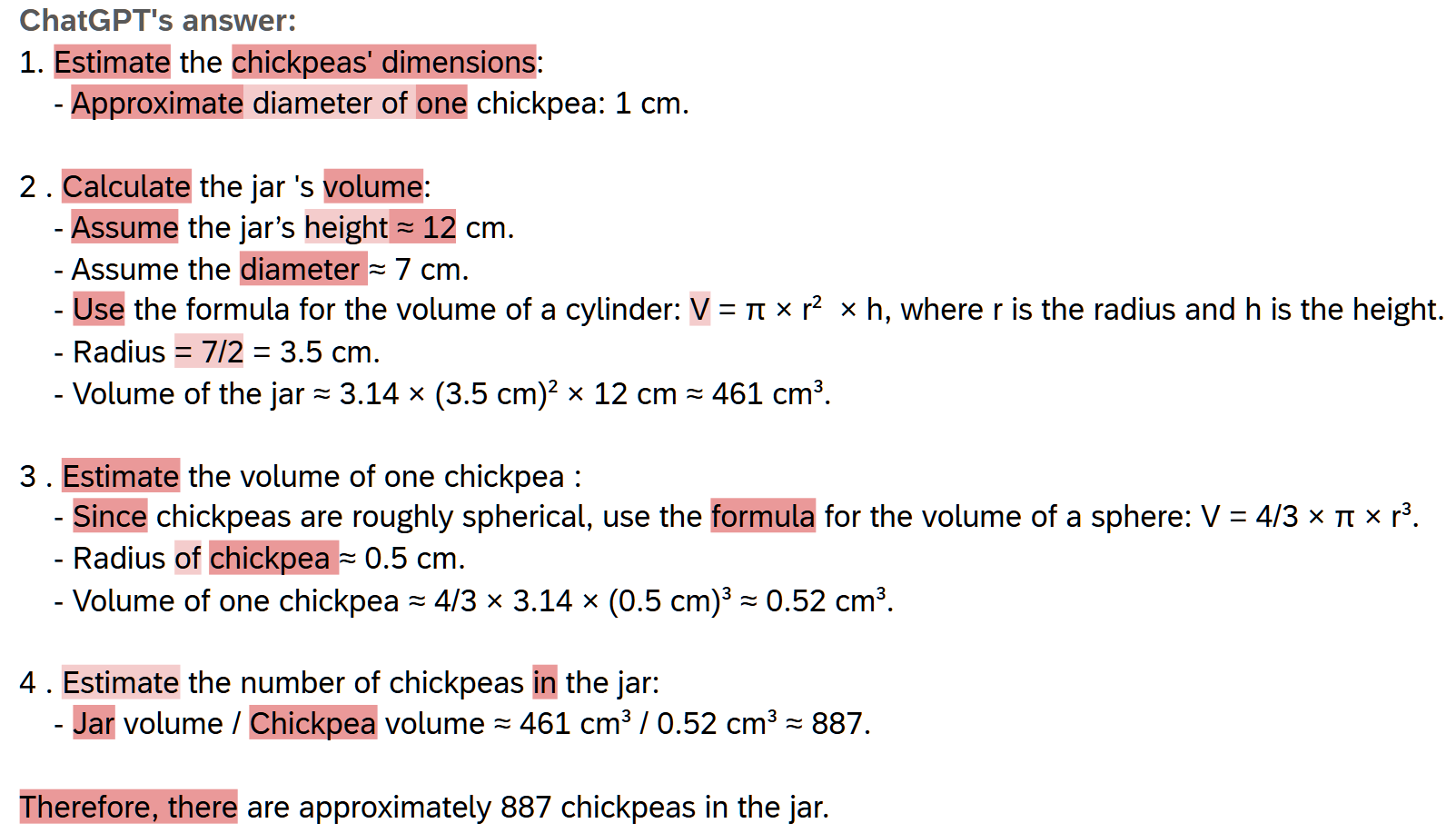}
        \label{subfig:ne_advice_hl}
        \caption{\hl}
    \end{subfigure}
    \begin{subfigure}[t]{0.49\linewidth}
        \raggedright
        \includegraphics[scale=0.23]{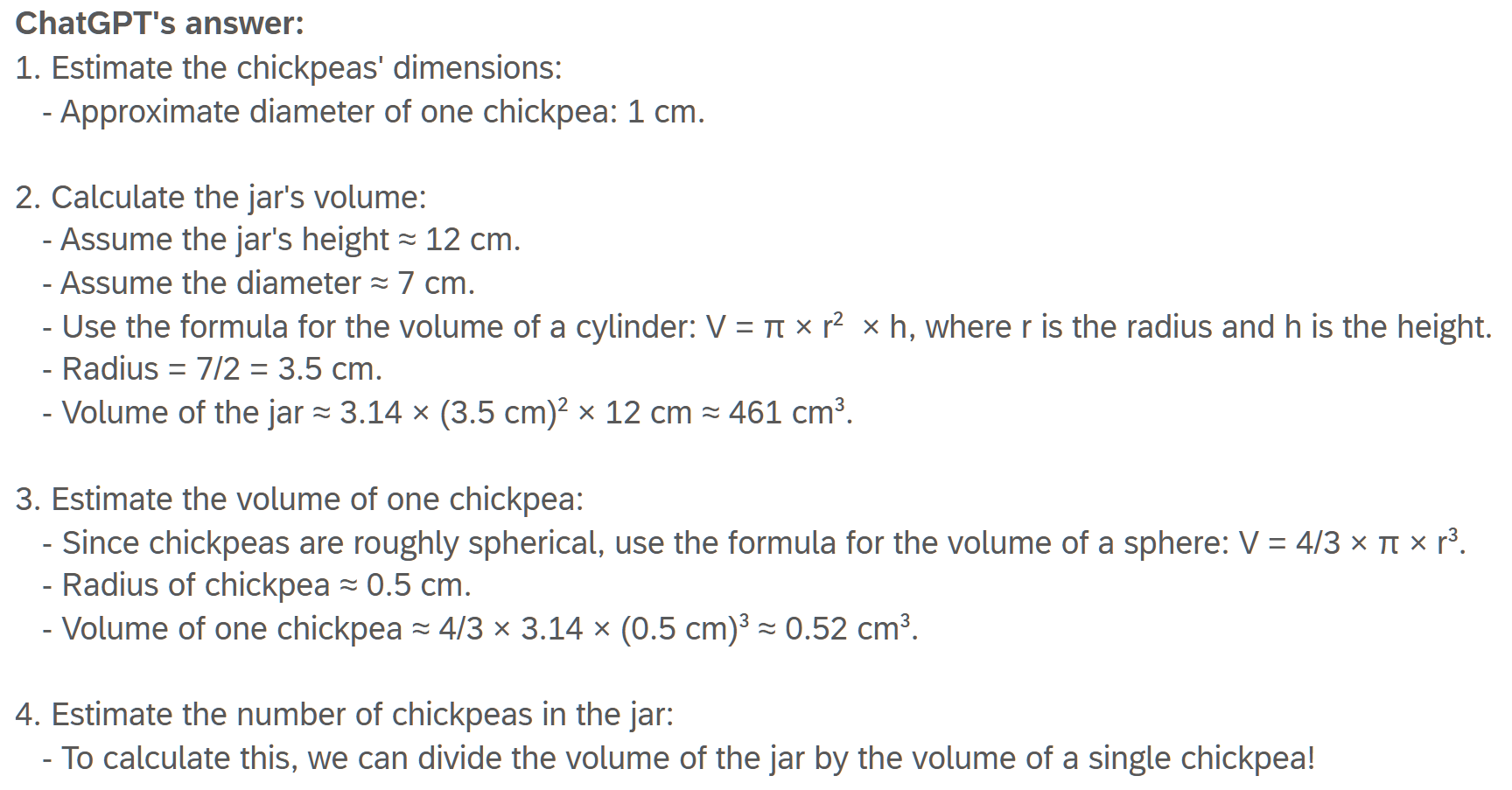}
        \label{subfig:ne_advice_expl}
        \caption{\implicit}
    \end{subfigure} 
    \caption{Sample LLM advice for all four conditions for an Numerical Estimation question.}
    \label{fig:ne_advice}
\end{figure*}

\section{Levels of Advice in Numerical Estimation}
\label{app:five_levels}
\begin{figure*}[h!]
    \centering
    \begin{subfigure}[t]{\linewidth}
        \centering
        \includegraphics[width=\linewidth]{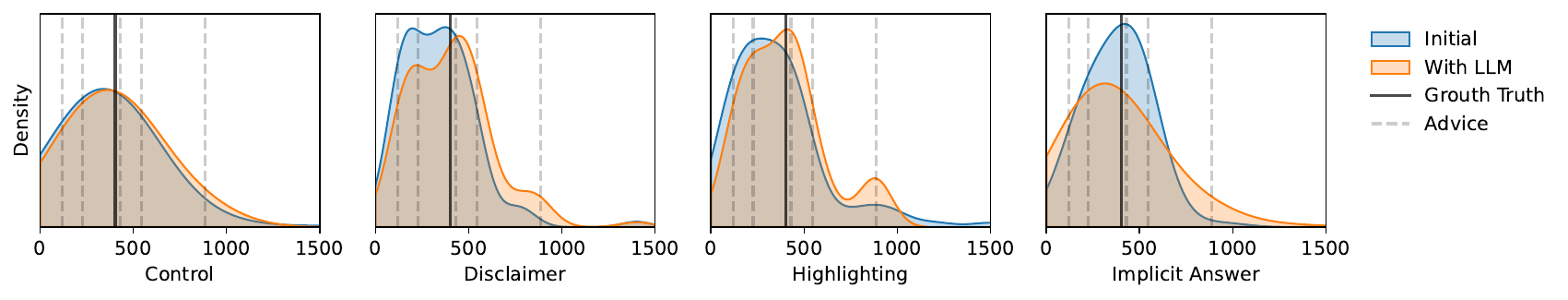}
        \caption{Chickpea estimation task.}
        \label{fig:chickpea_levels}
    \end{subfigure}%
    \\
    \begin{subfigure}[t]{\linewidth}
        \centering
        \includegraphics[width=\linewidth]{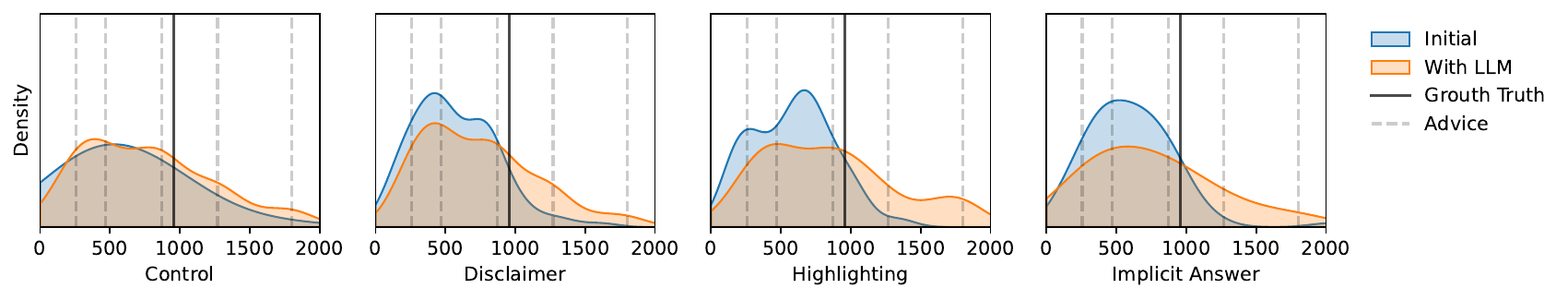}
        \caption{Jelly bean estimation task.}
        \label{fig:jelly_levels}
    \end{subfigure}
    \caption{\edit{Distribution of the participants' initial (blue) and final (orange) estimations for both Numerical Estimation tasks. The true count is indicated by the black line. The five gray dotted lines mark the different levels of LLM advice that were randomly provided to the participants.}}
    \label{fig:five_levels}
\end{figure*}

\edit{The five levels of LLM advice provided for the two questions in the Numerical Estimation task are illustrated in Figure \ref{fig:five_levels}, alongside the distribution of the participants' estimates before and after receiving the advice.}

\section{Confidence Analysis for Other WoA Thresholds}
\label{app:woa_conf}
\begin{figure*}[h!]
    \centering
    \begin{subfigure}[t]{0.4\linewidth}
        \centering
        \includegraphics[width=\linewidth]{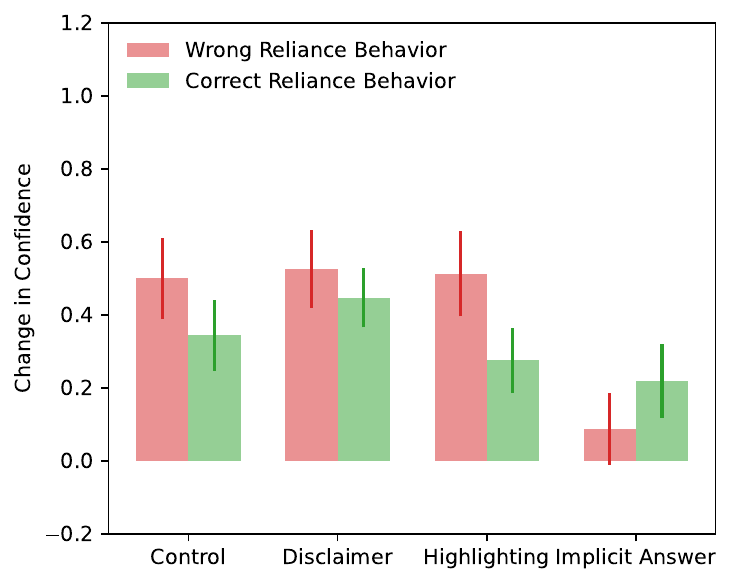}
        \caption{Correct vs. wrong reliance behaviours.}
        \label{fig:NE_reliance_conf_0.01}
    \end{subfigure}%
    ~ 
    \hspace{2em}
    \begin{subfigure}[t]{0.4\linewidth}
        \centering
        \includegraphics[width=\linewidth]{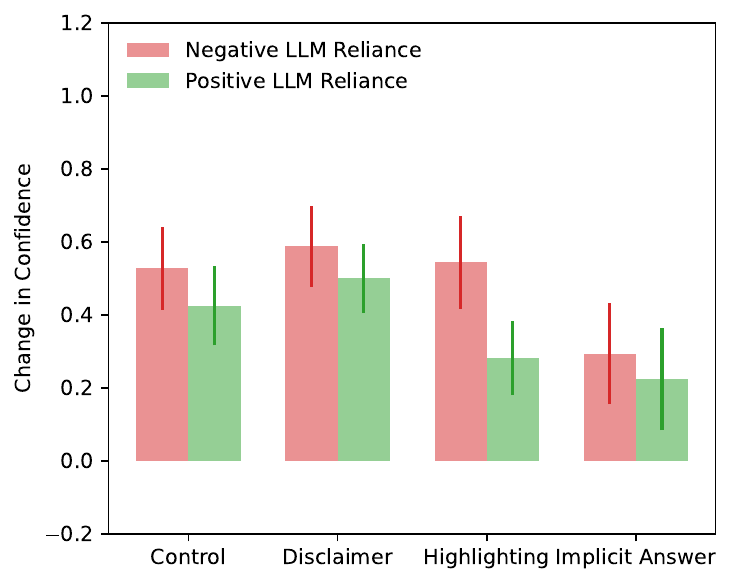}
        \caption{Appropriate vs. over-reliance on LLM advice.}
        \label{fig:NE_TP_FP_conf_0.01}
    \end{subfigure}
    \caption{Change in confidence at the question level for Numerical Estimation with a WoA threshold of 0.01 (highest sensitivity to reliance).}
    \label{fig:NE_conf_0.01}
\end{figure*}

\begin{figure*}[h!]
    \centering
    \begin{subfigure}[t]{0.4\linewidth}
        \centering
        \includegraphics[width=\linewidth]{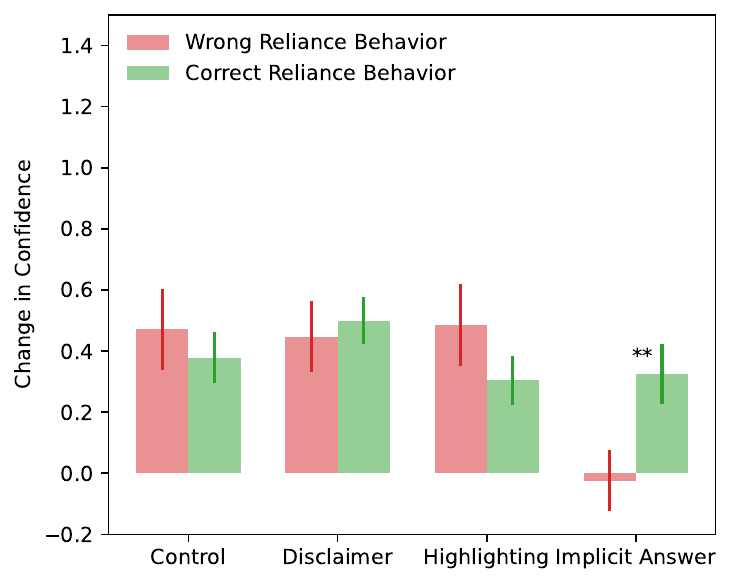}
        \caption{Correct vs. wrong reliance behaviours.}
        \label{fig:NE_reliance_conf_0.99}
    \end{subfigure}%
    ~ 
    \hspace{2em}
    \begin{subfigure}[t]{0.4\linewidth}
        \centering
        \includegraphics[width=\linewidth]{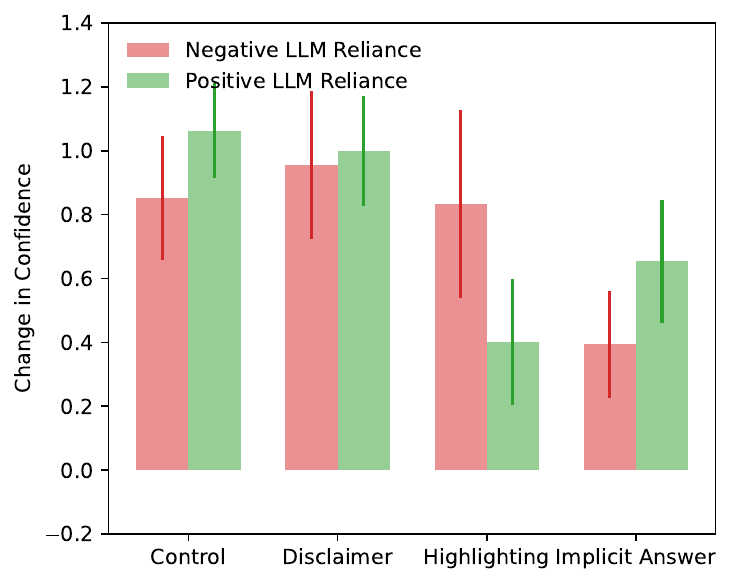}
        \caption{Appropriate vs. over-reliance on LLM advice.}
        \label{fig:NE_TP_FP_conf_0.99}
    \end{subfigure}
    \caption{Change in confidence at the question level for Numerical Estimation with a WoA threshold of 0.99 (lowest sensitivity to reliance).}
    \label{fig:NE_conf_0.99}
\end{figure*}

The confidence calibration graphs for Numerical Estimation at the other WoA thresholds (0.01 and 0.99) are listed here as Figure \ref{fig:NE_conf_0.01} and \ref{fig:NE_conf_0.99}.

\section{\edit{Effect of Prior LLM Experience}}
\label{app:usage}
\edit{
As an additional analysis, we examine whether participants' baseline frequencies of using LLM tools affected their behaviours and perceptions. The participants reported their prior experience with LLMs in the survey as \textit{never tried}, \textit{almost never but have tried}, \textit{monthly}, \textit{weekly}, and \textit{daily}, which we map into ordinal values from 0 to 4. When controlling for condition, a partial Spearman correlation reveals no significant correlations between LLM usage and reliance on the advice for either task, LSAT ($r=.02, p=.5$ for a binary reliance variable) or Numerical Estimation ($r=.06, p=.1$ for the continuous WoA variable). Nor are there correlations with the changes in confidence for LSAT  ($r\approx0, p=.9$) and Numerical Estimation ($r=.03, p=.4$). However, there is significant positive association between frequency of usage and all three categories of post-task perception: confidence ($r=.12, p<.05$), accuracy ($r=.14, p<.01$), and helpfulness ($r=.13, p<.01$). This suggests that while people who were more familiar and open to using LLM tools generally hold higher view of ChatGPT's advice (across all conditions), it did not impact actual rates of reliance in the task.
}

\end{document}